\documentclass[11pt, epsfig]{article} 
\usepackage{graphicx}
  \usepackage{amsthm,amsfonts}
  \usepackage{amsmath}
  \usepackage{color}
\newcommand{\bea}   {\begin{eqnarray}}
\newcommand{\eea}   {\end{eqnarray}}

\begin{document}
\renewcommand{\thefootnote}{\fnsymbol{footnote}}

\thispagestyle{empty}

\title{${\mathbb Z}_2\times {\mathbb Z}_2$-graded Lie Symmetries of the L\'evy-Leblond Equations}
\author{N. Aizawa\thanks{{E-mail: {\em aizawa@p.s.osakafu-u.ac.jp}}},\quad 
Z. Kuznetsova\thanks{{E-mail: {\em zhanna.kuznetsova@ufabc.edu.br}}},\quad
H. Tanaka\thanks{{E-mail: {\em s\_h.tanaka@p.s.osakafu-u.ac.jp}}}
\quad and\quad F.
Toppan\thanks{{E-mail: {\em toppan@cbpf.br}}}
\\
\\
}
\maketitle

\centerline{$^{\ast \ddag}$
{\it Department of Physical Science, Graduate School of Science,}}
{\centerline {\it\quad
Osaka Prefecture University, Nakamozu Campus,}}
{\centerline{\it\quad Sakai, Osaka 599-8531 Japan.}}
\centerline{$^{\dag}$
{\it UFABC, Av. dos Estados 5001, Bangu,}}{\centerline {\it\quad
cep 09210-580, Santo Andr\'e (SP), Brazil.}
\centerline{$^{\S}$
{\it CBPF, Rua Dr. Xavier Sigaud 150, Urca,}}{\centerline {\it\quad
cep 22290-180, Rio de Janeiro (RJ), Brazil.}
~\\
\maketitle
\begin{abstract} 

The first-order differential L\'evy-Leblond equations (LLE's) are the non-relativistic analogs of the Dirac equation, being square roots of ($1+d$)-dimensional Schr\"odinger or heat equations. Just like the Dirac equation, the LLE's possess a natural supersymmetry. In previous works it was shown that non supersymmetric PDE's (notably,  the 
Schr\"odinger equations for free particles or in the presence of a harmonic potential), admit a natural ${\mathbb Z}_2$-graded Lie  symmetry.\par
In this paper we show that, for a certain class of supersymmetric PDE's, a natural ${\mathbb Z}_2\times{\mathbb Z}_2$-graded Lie  symmetry appears. In particular, we exhaustively investigate the symmetries of the $(1+1)$-dimensional L\'evy-Leblond Equations, both in the free case and for the harmonic potential. In the free case a ${\mathbb Z}_2\times{\mathbb Z}_2$-graded Lie superalgebra, realized by first and second-order differential symmetry operators, is found. In the presence of a non-vanishing quadratic potential, the Schr\"odinger invariance is maintained, while the ${\mathbb Z}_2$- and ${\mathbb Z}_2\times{\mathbb Z}_2$- graded extensions are no longer allowed. \par
The construction of the ${\mathbb Z}_2\times {\mathbb Z}_2$-graded Lie symmetry of the ($1+2$)-dimensional free heat LLE introduces a new feature, explaining the existence of first-order differential symmetry operators not entering the super Schr\"odinger algebra.
~\\\end{abstract}
\vfil

\rightline{CBPF-NF-004/16}

\newpage
\section{Introduction}

In this paper we prove that a mathematical structure introduced  roughly 40 years ago by physicists aiming at generalizing the notion of supersymmetry (see \cite{{rw1},{rw2},{lr},{sch}}) appears, highly unexpectedly, as a dynamical symmetry of a class of equations which were introduced approximately 50 years ago \cite{ll}. \par
The prototype of these systems, the $(1+3)$-dimensional L\'evy-Leblond equation of \cite{ll}, is quite important in its own. Indeed,  it is Galilei-invariant and admits a well-defined probability interpretation. It possesses a better gyromagnetic ratio ($=2$) than the Pauli equation, incorporating in a natural way the spin in a non-relativistic setting. The original equation and several of its variants (not only in ($1+3$)-dimensions) have been investigated in several different physical contexts, see e.g. \cite{{gn},{ggt},{hor},{dhp},{hpv}}. \par
That such an important equation admits a (so far unnoticed) hidden symmetry of new type is 
already remarkable. Even more important, this new type of hidden symmetry can possibly emerge in several relevant physical systems. Some key observations have to be made. The ${\mathbb Z}_2\times{\mathbb Z}_2$-graded structure of \cite{{rw1},{rw2},{lr},{sch}}}, not admitting a relativistic spin-statistics connection, is expected (if any) to be encountered in non-relativistic systems, or in dynamical systems possessing anyons or parastatistics. \par
Plenty of models are potential candidates for a ${\mathbb Z}_2\times {\mathbb Z}_2$-graded hidden symmetry. Among the main tools that we employed in this paper we have to mention Clifford algebras and nilpotent operators. A nilpotent operator is a necessary ingredient for both  the construction of L\'evy-Leblond equations and the introduction of topological theories based on the \cite{wit} cohomological framework.  One particular appealing application of Clifford algebras consists in encoding emergent Majorana fermions in condensed matter, see \cite{{kit1},{kit2}}. These theories are highly investigated in the context of topological quantum computation since they offer topological protection from quantum decoherence. In these class of theories the topology enters as a result of braid statistics, see \cite{kauf}. \par
Under Fermionization, which is the converse, dual property of Bosonization \cite{col}, fermionic degrees of freedom emerge in purely bosonic theories from non-local bosonic operators. It is quite noticeable that emergent supersymmetry appears in mechanical models of masses connected by springs and mimicking the properties of topological insulators \cite{kalu}, or even in topological metamaterials (where the springs are replaced by rigid bars, see \cite{{vit1},{vit2}}). There is lot of evidence by now that emergent supersymmetry (based on a single ${\mathbb Z}_2$-gradation) is a reality which helps characterizing and deriving properties of the associated models. It is therefore natural to expect that an extra
${\mathbb Z}_2$-gradation (giving a full ${\mathbb Z}_2\times {\mathbb Z}_2$-graded structure) could also emerge and be recognized to exist in theories beyond the L\'evy-Leblond generalized models.\par It is clear that, each time a new class of symmetry is found, it offers the possibility of unexpected non-trivial constraint (for instance in the spontaneous breaking or Higgs mechanism) which could not be envisaged without the knowledge of the new type of symmetry.  In this paper the presence of the ${\mathbb Z}_2\times {\mathbb Z}_2$-graded symmetry allows us to identify new symmetry operators. We should also mention some preliminary evidence, based on specific models under current investigation, that ${\mathbb Z}_2\times {\mathbb Z}_2$-graded symmetry, employed as a spectrum-generating algebra, could allow derive the full spectrum of a theory from a single irreducible lowest weight representation. The superalgebra (contained as subalgebra inside the ${\mathbb Z}_2\times {\mathbb Z}_2$-structure) only possesses partial information presenting the spectrum as the decomposition into a direct sum of (several) of its irreducible representations.\par
 The above considerations make clear the potential range of applications of the ${\mathbb Z}_2\times {\mathbb Z}_2$-graded symmetry in dynamical systems. \par
We are now focusing on the main technical results of our work.
In this paper we prove the existence of a finite ${\mathbb Z}_2\times {\mathbb Z}_2$-graded Lie symmetry, realized by first-order and second-order differential operators, of the 
free generalized L\'evy-Leblond equations (the ($1+1$)- and ($1+2$)-dimensional cases are here explicitly discussed).\par
It was pointed out in \cite{top} that finite Lie superalgebras can be symmetries for a certain class of purely bosonic  partial differential equations  (including the cases of the free particle and of the harmonic oscillator)\footnote{The fact that a Lie superalgebra appears even in a purely bosonic setting is not so surprising. Indeed, for the harmonic oscillator, the old results of \cite{wig} can be expressed, in modern language, by stating that the Fock vacuum of creation/annihilation operators can be replaced by a lowest weight representation of an $osp(1|2)$ spectrum-generating superalgebra.}.
The recognition that a symmetry superalgebra is present was later applied \cite{akt} in the context of Conformal Galilei Algebras to identify new bosonic invariant partial differential equations.\par
It is therefore natural to pose the question: what happens in the case of a supersymmetric system, namely 
one which already possesses a ${\mathbb Z}_2$-graded structure?  Is, in that case, a second ${\mathbb Z}_2$-gradation present? To give an answer we started investigating the generalized L\'evy-Leblond equations, which are the non-relativistic analogs of the Dirac equations, being the square roots of the heat or of the Schr\"odinger equation (with or without potential terms). The original L\'evy-Leblond equation \cite{ll}
is the non-relativistic wave equation of a spin-$\frac{1}{2}$ particle in the ordinary $(1+3)$-dimensional space-time and possesses a natural supersymmetry. In Section {\bf 2} we discuss at length the generalized L\'evy-Leblond equations, induced by first-order matrix differential operators, and their construction from the relevant  Clifford algebras.\par
This investigation about graded symmetries requires a preliminary understanding of the ${\mathbb Z}_2\times {\mathbb Z}_2$-graded Lie
(super)algebra structures. Generalizations of Lie and super-Lie algebras were introduced  and named
 {\em color} algebras (for certain resemblances to parastatistics) in \cite{rw1,rw2}. They were further investigated in \cite{lr,sch}. Nowadays there is a small body of literature about these structures, dealing with
possible physical applications (see, e.g., \cite{{vas},{jyw},{zhe},{tol}}) and a larger number of works devoted to their mathematical development (for more recent papers see, e.g., \cite{{ps},{cov}} and references therein). The absence of a spin-statistics connection in the relativistic setting (since we are working in a non-relativistic setting, this is not a concern for us) has been the major impediment for the full development of the topic of generalized supersymmetries. To make this paper self-contained, the relevant properties of ${\mathbb Z}_2\times {\mathbb Z}_2$-graded color Lie (super)algebras are presented in an Appendix.\par
Our investigation requires the further notion, following \cite{fuzh}, of {\em symmetry operator}. We consider two classes of symmetries operators; they can be recovered either from commutators or  from anticommutators,
see equations (\ref{comm}) and, respectively, (\ref{anticomm}). Quite interestingly, we need symmetry operators belonging to both classes in order to produce a ${\mathbb Z}_2\times {\mathbb Z}_2$-graded symmetry.\par
In this work we computed the complete list of symmetry operators for the L\'evy-Leblond square root of the free heat equation in $(1+1)$-dimensions (a $2\times 2$ matrix operator) and the square root of the free Schr\"odinger equation in $(1+1)$-dimensions (a $4\times 4$ matrix operator in the real counting). We also computed the symmetry operators recovered from commutators for the L\'evy-Leblond square root of the heat equation with quadratic potential (a $4\times 4$ matrix operator). Once identified the symmetry operators, we investigated the closed, finite (graded) Lie symmetry algebras induced by them. We proved, in particular, that the L\'evy-Leblond square roots of the $(1+1)$-dimensional free heat and free Schr\"odinger 
equations possess a super Schr\"odinger symmetry algebra with maximal ${\cal N}=1$ extension (super Schr\"odinger algebras were discussed in \cite{{ggt},{dh}}; a Kaluza-Klein derivation of the super Schr\"odinger algebra of the L\'evy-Leblond equation was given in \cite{hor}).  The
${\mathbb Z}_2\times {\mathbb Z}_2$-graded Lie symmetry  ${\cal G}_{{\mathbb Z}_2\times{\mathbb Z}_2}
$ of first and second-order differential operators,
spanned by the $13$ generators in (\ref{z2z2sym}), is found.\par
The situation is quite different for the square root of the heat equation with quadratic potential. A Schr\"odinger symmetry algebra is still present. We proved that it can not, on the other hand, be extended to a super Schr\"odinger symmetry algebra. \par
In Appendix {\bf B} we present the construction of the ${\mathbb Z}_2\times {\mathbb Z}_2$-graded Lie symmetry superalgebra of the L\'evy-Leblond operator associated with the ($1+2$)-dimensional free heat equation. A new feature emerges. The ${\mathbb Z}_2\times {\mathbb Z}_2$-graded symmetry includes first-order differential symmetry operators which do not belong to the two-dimensional super Schr\"odinger algebra.\par We postpone to the Conclusions a more detailed summary of our results, with comments and a discussion of their implications.\par
The scheme of the paper is as follows. In Section {\bf 2} we introduce the generalized L\'evy-Leblond operators and their relation to Clifford algebras. In Section {\bf 3} we introduce, following \cite{fuzh}, the notion of {\em symmetry operators}. In Section {\bf 4} the full list  of symmetry operators of the 
L\'evy-Leblond square roots of the free heat and free Schr\"odinger equation in $(1+1)$-dimensions is
presented. Finite (graded) Lie algebras induced by these symmetry operators (including the ${\cal N}=1$ super Schr\"odinger algebra and the ${\mathbb Z}_2\times {\mathbb Z}_2$-graded symmetry) are presented in Section {\bf 5}. Section {\bf 6} is devoted to the symmetry of the L\'evy-Leblond square root of the $(1+1)$-dimensional heat equation with quartic potential. A summary of our results is given in the Conclusions. A self-contained presentation of the ${\mathbb Z}_2\times {\mathbb Z}_2$-graded color Lie (super)algebras (and their relation to quaternions and split-quaternions) is given in Appendix {\bf A}. We present in Appendix {\bf B} the construction of the ${\mathbb Z}_2\times{\mathbb Z}_2$-graded symmetry of the free ($1+2$)-dimensional L\'evy-Leblond equation. In Appendix {\bf C} we discuss the possibility of introducing different ${\mathbb Z}_2\times {\mathbb Z}_2$-graded
Lie (super)algebras, based on different assignments of gradings to given operators.

\section{On Clifford algebras and generalized L\'evy-Leblond equations }\par

The original L\'evy-Leblond equation \cite{ll} is the square root of the Schr\"odinger equation in $1+3$ dimensions. Generalized L\'evy-Leblond equations are square roots of heat or Schr\"odinger equations in an arbitrary number of space dimensions; they can be systematically constructed from Clifford algebras irreducible representations. We introduce here the basic ingredients for this general scheme, focusing on issues such as complex structure, introduction of a potential term and so on. \par
We remind that the irreducible representations (over ${\mathbb R}$) of the $Cl(p,q)$ Clifford algebras (the enveloping algebras whose gamma-matrix generators satisfy $\gamma_i\gamma_j+\gamma_j\gamma_i= 2\eta_{ij}$, where $\eta_{ij}$ is a diagonal matrix with $p$ positive ($+1$) and $ q$ negative ($-1$) entries) 
can be obtained by tensoring four $2\times 2$ matrices, see e.g. \cite{oku}. It is convenient to follow the presentation of \cite{tv}.
The $2\times 2$ matrices can be identified with (four) letters. General gamma matrices can be expressed, since no ambiguity arises, as words in a $4$-letter alphabet by dropping the symbol of tensor product
``$\otimes$". By expressing
\bea\label{IXYA}
&I = \left(\begin{array}{cc} 1&0\\0&1\end{array}\right),\quad
X = \left(\begin{array}{cc} 1&0\\0&-1\end{array}\right),\quad
Y = \left(\begin{array}{cc} 0&1\\1&0\end{array}\right),\quad
A = \left(\begin{array}{cc} 0&1\\-1&0\end{array}\right),&
\eea
the split-quaternions, see Appendix {\bf A}, can be represented as
\bea
&{\widetilde e}_0= I,\quad
{\widetilde e}_1 = Y,\quad
{\widetilde e}_2 = X,\quad
{\widetilde e}_3 = A,&
\eea
where $X,Y,A$ are the gamma matrices realizing the $Cl(2,1)$ Clifford algebra.\par
The quaternions can be realized as $4\times 4$ real matrices. With the adopted  convention of dropping the tensor product symbol they can be presented as
\bea
&{ e}_0= II,\quad
{e}_1 = AI,\quad
{e}_2 = XA,\quad
{e}_3 = YA.&
\eea
Some comments are in order:\\
{\em i}) a matrix is block-diagonal if, in its associated word, the first letter is either $I$ or $X$. Conversely, it is block-antidiagonal if the first letter is $Y$ or $A$,\\
{\em ii}) a matrix is (anti)symmetric, depending on the number (even or odd) of $A$'s in its word,\\
{\em iii})  in real form the complex structure is defined by a real matrix $J$ such that $J^2=-{\mathbb I}$.\\ The complex-structure
preserving matrices commute with $J$. If the complex structure is preserved, complex numbers can be used. A possible choice for the $4\times 4$ real matrices complex structure is setting $J=IA$, since $(IA)^2=-II=-{\mathbb I}_4$. Therefore, the eight $4\times 4$ complex-structure preserving matrices are  $II, XI, YI, AA, IA, XA, YA, AI$.\par
Depending on the $(p,q)$ signature, the maximal number $p+q$ of Clifford gamma-matrix generators that can be accommodated in $2^n\times 2^n$ real matrices is
\bea
2\times 2 & :&   Cl(2,1),\nonumber\\
4\times 4 &:& Cl(3,2),\quad Cl(0,3),\nonumber\\
8\times 8 &:& Cl(4,3),\quad Cl(5,0),\quad Cl(1,4),\quad Cl(0,7),
\eea
and so on.\par
Clifford algebras can be used to introduce Supersymmetric Quantum Mechanics (SQM). In its simplest version
(the one-dimensional, ${\cal N}=2$ supersymmetry, with $x$ as a space coordinate) the two supersymmetry operators    
$Q_1, Q_2$ need to be block anti-diagonal, hermitian and complex structure preserving first-order real differential operators. Moreover, they have to satisfy  the ${\cal N}=2$ SQM algebra
\bea
\{Q_i, Q_j\} = 2\delta_{ij} H, \quad &&\quad [H, Q_i]=0,
\eea
where $H$ is the Hamiltonian.\par
The irreducible representation (in real counting) requires $4\times 4$ matrices. By setting the complex structure to be $J=IA$, the most general solution, for an arbitrary function $f(x)$ (the prepotential), can be expressed as
\bea
Q_1 &=& AI\partial_x + YI f(x),\nonumber\\
Q_2 &=& YA\partial_x +AA f(x),\nonumber\\
H &=& II( -\partial_x^2 +f(x)^2) + XI f_x(x)= \mathbb{I}_4 (-\partial_x^2+f(x)^2) +N_f f_x(x).
\eea
The hamiltonian $H$ is diagonal. In its upper block the potential is $V_+(x) = f(x)^2+f_x(x)$, while in its lower block the potential is $V_-(x) = f(x)^2-f_x(x)$.\par
$XI$ defines the fermion number operator $N_f$ ($N_f=XI$). All operators act on two real component bosonic and two real component fermionic fields. For energy eigenvalues $E>0$ the $H_\pm = -\partial_x^2+ f(x)^2\pm f_x(x) $ Hamiltonians share the same spectrum. The free case is obtained by taking $f(x)=0$; the harmonic oscillator case is recovered by taking $f(x)=cx$.\par
Since $Q_1,Q_2,H$ commute with $J$ and preserve the complex structure, they can also be expressed as $2\times 2$ {\em complex} matrices.
\par
The construction of a (generalized) L\'evy-Leblond real differential operator requires the linear combination of two gamma matrices (one with a positive and the other one with a negative square) in order to produce a nilpotent operator which eliminates the $\partial_t^2$ term present in the relativistic Klein-Gordon equation. \par
The original L\'evy-Leblond equation (the square root of the Schr\"odinger equation in $1+3$ dimensions), can be recovered from  the $Cl(1,4)$ Clifford algebra, whose matrices act on $8$ real component fields. 
Since $Cl(1,4)$ has a quaternionic (and therefore, {\em a fortiori}, complex) structure \cite{oku}, the $8\times 8$ real matrices preserving the complex structure can be described as $4\times 4$ complex matrices acting on a multiplet of $4$ component complex fields.\\
We call a  first-order real matrix differential operator a ``generalized L\'evy-Leblond operator" if
it is the square root of either the heat or the Schr\"odinger equation in $1+d$ dimensions (with or without a potential term). Further properties can be imposed. It could be required the operator to be
block anti-diagonal and anticommuting with the fermion number operator, so that it mutually exchanges bosons into fermions.\par
The minimal matrix size to accommodate a L\'evy-Leblond square root of the $1+1$ heat equation is $2$.
Indeed, we can set
\bea
{\overline \Omega}\Psi(x,t)=0,\quad {\overline\Omega} = \frac{1}{2}(A+Y)\partial_t+\frac{1}{2}(A-Y)\lambda + X\partial_x  &\rightarrow & {\overline \Omega}^2={\mathbb I}_2 (-\lambda\partial_t+\partial_x^2),\eea
with $\lambda$ an arbitrary real number.\par
One should note that ${\overline \Omega}$ is neither block-antidiagonal nor complex structure preserving. 
The minimal solution to have a block antidiagonal (and complex-structure preserving) operator requires $4\times 4$ real matrices. A convenient basis for the five $4\times 4$ $Cl(3,2)$ gamma matrices is given by
$AA, AX, AY, XI, YI$. The complex structure can be defined by $J=AI$. The three complex-structure preserving matrices $AA, AX, AY$ satisfy the $p=1$, $q=2$ gamma-matrix relations. \par
A block antidiagonal, complex structure preserving, square root of the free heat equation in $1+1$ dimensions is given by
\bea\label{heatfree}
\Omega_{heat,free} &=& \frac{1}{2}(AA+AY)\partial_t +\frac{1}{2}(AA-AY)\lambda +AX\partial_x,\nonumber\\
\Omega_{heat,free}^2 &=& II(\lambda\partial_t-\partial_x^2)= {\mathbb I}_4 (\lambda\partial_t-\partial_x^2).
\eea
The introduction of a potential term requires the use of the extra antidiagonal gamma matrix $YI$. Therefore,
in the presence of a non-vanishing potential, the $4\times 4$ L\'evy-Leblond operator  cannot preserve the  complex-structure. We have
\bea\label{heat}
\Omega_{heat}&=&  \frac{1}{2}(AA+AY)\partial_t +\frac{1}{2}(AA-AY)\lambda +AX\partial_x +YI f(x),\nonumber\\
\Omega_{heat}^2&=& II(\lambda\partial_t-\partial_x^2+f(x)^2)+XXf_x(x).
\eea
Due to the mentioned property, $\Omega_{heatfree}$ in (\ref{heatfree}) can be represented by $2\times 2$ complex matrices, while this is not true for $\Omega_{heat}$ in (\ref{heat}) if $f(x)$ is non-vanishing.\par
The (free) Schr\"odinger equation requires the complex structure.  The square root of the free Schr\"odinger equation can be obtained by modifying $\Omega_{heat,free}$ so that 
\bea
\Omega_{Sch,free} &=& \frac{1}{2}(AA+AY)\cdot AI\partial_t +\frac{1}{2}(AA-AY)\lambda +AX\partial_x,
\eea
where ``$\cdot$" denotes the ordinary matrix multiplication.\par
Its square gives
\bea
\Omega_{Sch,free}^2 &=& AI\lambda\partial_t-II\partial_x^2,
\eea
namely a Schr\"odinger equation on $4$-component real fields (which can be equivalently expressed as an equation for $2$-component complex fields).\par
The introduction of a non-vanishing potential for the $1+1$ Schr\"odinger equation requires a L\'evy-Leblond operator $\Omega_{Sch}$ acting on at least $8$ real component fields. With the adopted conventions $\Omega_{Sch}$ can be expressed as
\bea
\Omega_{Sch} &=& \frac{1}{2}(AAA+AAY)\cdot  AII\partial_t +\frac{1}{2}(AAA-AAY)\lambda +AYI\partial_x+AAX f(x),\nonumber\\
\Omega_{Sch}^2 &=& -AII\lambda\partial_t+III(-\partial_x^2+f(x)^2) +IXXf_x(x),
\eea
the complex structure being defined by $J=AII$.\par
It is straightforward to systematically construct, following this scheme and with the tools in \cite{tv}, generalized L\'evy-Leblond operators in ($1+d$) dimensions.

\section{On symmetries of matrix partial differential equations}

Our investigation heavily relies on the notion of {\em symmetry operator} defined in \cite{fuzh}. We recall that
a symmetry operator $Z$ of a matrix partial differential equation induced by an operator $\Omega$ maps a given solution $\Psi(\vec{x})$ 
into another solution $Z\Psi(\vec{x})$, according to
\bea
\Omega\Psi({\vec{x}}) =0 &\Rightarrow & \Omega(Z\Psi({\vec{x}}))|_{\Omega\Psi=0} =0.
\eea
$Z$ can be any kind of operator. In the traditional Lie viewpoint the symmetry group of a differential equation is generated by a subset of operators which are closed under commutation. This condition is relaxed for superalgebras or ${\mathbb Z}_2\times{\mathbb Z}_2$ graded Lie algebras. Based on the grading of the symmetry operators, the closure requires both commutators and anticommutators. 
\par
If we restrict $Z$ to be a differential operator of finite order, $Z$ can be called a {\em symmetry operator} \cite{fuzh} if
the following sufficient condition for symmetry is fulfilled:
\par 
either
\bea\label{comm}
\relax [\Omega, Z] &=& \Phi_Z({\vec x})\Omega,
\eea

or
\bea\label{anticomm}
\{\Omega, Z\}&= \Phi_Z({\vec{x}})\Omega,
\eea
where $\Phi_Z({\vec{x}})$ is a $n\times n$ matrix-valued function of the ${\vec{x}}$ space(time) coordinates.\par
If there is no ambiguity (certain operators, like the identity ${\mathbb I}$, satisfy both (\ref{comm}) and (\ref{anticomm})) throughout the text we denote with $\Sigma$'s the symmetry operators satisfying (\ref{comm}) and with $\Lambda$'s the symmetry operators satisfying (\ref{anticomm}).\par
In our applications we consider, initially, first-order differential symmetry operators. Second-order differentials
symmetry operators are also constructed by taking suitable anticommutators of the previous operators.\par
It is important to note that, if $\Sigma_1,\Sigma_2$ are two operators satisfying (\ref{comm}), then by construction the identity 
\bea
[\Omega, [\Sigma_1,\Sigma_2]]&=& \Phi_{[\Sigma_1,\Sigma_2]}\Omega
\eea
is satisfied. This does not imply, however, that the commutator $[\Sigma_1,\Sigma_2]$ is a symmetry operator according to the given definition. Indeed, $\Phi_{[\Sigma_1,\Sigma_2]}$ can be a differential operator and not necessarily a matrix-valued function. In the following, see e.g. (\ref{omegatomega},\ref{symomegatomega}), some examples are given. A Lie algebra of symmetry operators requires $\Phi_{[\Sigma_1,\Sigma_2]}$ to be a matrix-valued function.

\section{Symmetries of the L\'evy-Leblond square root of the free heat and Schr\"odinger equations}

As discussed in Section {\bf 2}, the minimal realization of the L\'evy-Leblond square root of the $1+1$-dimensional free heat (Schr\"odinger) equation requires $2\times 2$ (and, respectively, $4\times 4$) matrices.\par
We produce here the exhaustive list of its symmetry operators.\par
With respect to Section {\bf 2} we slightly change the notations in order to easily extrapolate the results from the heat to the Schr\"odinger equation. It is convenient to set, as $2\times 2$ matrices,
\bea
\gamma_\pm=\frac{1}{2}(Y\pm  A) &,& \gamma_3= X,
\eea
where $X,Y,A$ and the identity $I\equiv {\mathbb I}$ are introduced in (\ref{IXYA}). The following relations are satisfied
\bea\label{gammas}
&\gamma_\pm^2=0, \quad \gamma_3^2={\mathbb I},\quad \gamma_\pm\gamma_\mp=\frac{1}{2}({\mathbb I}\pm \gamma_3),
\quad \gamma_3\gamma_\pm = \pm \gamma_\pm=-\gamma_\pm \gamma_3.&
\eea
Our analysis of the symmetries only uses the (\ref{gammas}) relations; therefore, the results below are representation-independent and can be extended to other representations (in particular to the case of $4\times 4$ matrices). For $4\times 4$ matrices we can introduce, following the convention of Section {\bf 2}, the $Cl(1,2)$ Clifford algebra generators
\bea
&\widetilde{\gamma}_1= AA,\quad
{\widetilde{\gamma}}_2 = AY,\quad
{\widetilde {\gamma}}_3 = AX.&
\eea
They define the complex structure $J$ given by
\bea
J&=&{\widetilde\gamma}_1{\widetilde\gamma}_2{\widetilde\gamma}_3=-AI, \quad\quad (J^2=-II=-{\mathbb I}).
\eea
A realization of the (\ref{gammas}) algebra in terms of $4\times 4$ matrices is given by setting
\bea\label{newgammas}
&\gamma_\pm= \pm\frac{1}{2}J\cdot({\widetilde\gamma}_1\pm{\widetilde\gamma}_2),\quad \gamma_3= J\cdot{\widetilde\gamma}_3.&
\eea
By construction $\gamma_\pm,\gamma_3$ defined in (\ref{newgammas}) commute with $J$ ($[J,\gamma_\pm]=[J,\gamma_3]=0$).\par
The L\'evy-Leblond operator $\Omega$ of the free heat equation is introduced (both in the $2\times 2$ and in the $4\times 4$ representations) as
\bea\label{omega}
\Omega &=& \gamma_+\partial_t +\gamma_-\lambda +\gamma_3\partial_x.
\eea
Its square is
\bea\label{omegasq}
\Omega^2&=& {\mathbb I}\cdot(\lambda\partial_t +\partial_x^2).
\eea
According to the convention introduced in Section {\bf 3} we denote with $\Sigma$'s the symmetry operators arising from commutators and with $\Lambda$'s the symmetry operators arising from anticommutators
\bea
\relax [\Sigma,\Omega] &=& \Phi_\Sigma\cdot \Omega, \label{sigma}\\
\{\Lambda,\Omega\}&=& \Phi_\Lambda\cdot\Omega,\label{lambda}
\eea
for some given matrix-valued function $\Phi_\Sigma$ or $\Phi_\Lambda$ in the $x,t$ variables.\par
The computation of the symmetry operators is tedious but straightforward. In the $2\times 2$ case the complete list (up to normalization) of first-order differential operators satisfying (\ref{sigma}) is given by
\bea\label{sigmasym}
H&=&{\mathbb I}\partial_t,\nonumber\\
D&=& -({\mathbb I}(t\partial_t+\frac{1}{2}x\partial_x+\frac{1}{2})+\frac{1}{4} \gamma_3),\nonumber\\
K&=& - ({\mathbb I}(t^2\partial_t+tx\partial_x-\frac{\lambda}{4}x^2+t)-\frac{1}{2}\gamma_+x +\frac{1}{2}\gamma_3t),\nonumber\\
P_+ &=& {\mathbb I}\partial_x,\nonumber\\
P_- &=&{\mathbb I}( t\partial_x-\frac{\lambda}{2}x) -\frac{1}{2}\gamma_+,\nonumber\\
C&=& {\mathbb I},\nonumber\\
\Omega_{z(x,t)}&=& z(x,t) \Omega = z(x,t)(\gamma_+\partial_t +\gamma_-\lambda +\gamma_3\partial_x).
\eea
The class of symmetry operators $\Omega_{z(x,t)}$ depends on an unconstrained function $z(x,t)$ ($\Omega$ is recovered by setting $z(x,t)=1$).\par
The complete set (up to normalization) of $2\times 2$ first-order differential operators satisfying 
(\ref{lambda}) is given by
\bea\label{lambdasym}
\Lambda_1&=& {\mathbb I},\nonumber\\
\Lambda_2&=& \gamma_+\partial_t-\lambda\gamma_-,\nonumber\\
\Lambda_3 &=& \gamma_+(t\partial_t+\frac{1}{2})-\gamma_-\lambda t +\frac{\lambda}{2}x{\mathbb I},\nonumber\\
\Lambda_4&=& \gamma_3\partial_t-2\gamma_-\partial_x,\nonumber\\
\Lambda_5&=& \gamma_3(t\partial_t+\frac{1}{4})+\gamma_-(-2t\partial_x+\frac{\lambda}{2})-\gamma_+\frac{x}{2}+\frac{1}{2}{\mathbb I},\nonumber\\
\Lambda_6&=& \gamma_3(\frac{1}{2}t^2\partial_t+\frac{t}{4})+\gamma_-(-t^2\partial_x+\frac{\lambda}{2}tx)+\gamma_+(-\frac{1}{2}tx\partial_t-\frac{x}{4})+(\frac{t}{2}-\frac{\lambda x^2}{8}){\mathbb I},\nonumber\\
{\widetilde \Lambda}_{w(x,t)}&=& w(x,t)(\gamma_+\partial_x-\frac{\lambda}{2}({\mathbb I}+\gamma_3)).
\eea
 The  ${\widetilde \Lambda}_{w(x,t)}$  class of operators depend on an unconstrained function $w(x,t)$.\par
The computation of the associated matrix-valued functions $\Phi_\Sigma(x,t), \Phi_\Lambda(x,t)$ is left as a simple exercise for the Reader.\par
As explained above, formulas (\ref{sigmasym}) and (\ref{lambdasym}) are representation-independent and provide symmetry operators for the $4\times 4$ matrix case as well. For this matrix representation, since the complex structure operator $J$ commutes with $\gamma_\pm,\gamma_3$, the number of symmetry operators is doubled with respect to the $2\times 2$ matrix representation. Indeed, for any given symmetry operator $\Sigma$ or $\Lambda$ entering (\ref{sigmasym}) or (\ref{lambdasym}), $J\cdot \Sigma$  ($J\cdot\Lambda$) is a symmetry operator satisfying (\ref{sigma}) (and, respectively, (\ref{lambda})). \par
We explicitly verified that no further symmetry operator is encountered, in the $4\times 4$ representation, besides the operators given in (\ref{sigmasym},\ref{lambdasym}) and their $J$-doubled counterparts. \par
The presence of the complex structure $J$ in the $4\times 4$ matrix case allows to induce, from the L\'evy-Leblond square root of the free heat equation, the L\'evy-Leblond square root of the free Schr\"odinger equation in $1+1$ dimensions. The most straightforward way consists in replacing, in the formulas above, $\lambda\mapsto\beta J$, for a real $\beta$. This substitution is allowed because $J$ commutes with all matrices entering (\ref{omega},\ref{sigmasym},\ref{lambdasym}).\par
We therefore get, for the $1+1$-dimensional free Schr\"odinger case,
\bea\label{omegabar}
{\overline\Omega} &=& \gamma_+\partial_t +\gamma_-\cdot J\beta +\gamma_3\partial_x,\nonumber\\
{\overline\Omega}^2&=& {\mathbb I}\cdot(J\beta\partial_t +\partial_x^2).
\eea

\section{The graded Lie symmetry algebras of the free equations}

We investigate here the properties of the closed graded Lie (super)algebras recovered from the symmetry operators entering (\ref{sigmasym}) and (\ref{lambdasym}).\par
We start by observing that the first six operators in (\ref{sigmasym}) span the one-dimensional Schr\"odinger algebra $sch(1)$. Their non-vanishing commutators are  
\bea\label{schr}
\relax [D,H]&=& H,\nonumber\\
\relax [D,K]&=& -K,\nonumber\\
\relax [H,K] &=& 2D,\nonumber\\
\relax [D,P_\pm]&=& \pm\frac{1}{2}P_\pm,\nonumber\\
\relax [H,P_-]&=& P_+,\nonumber\\
\relax [K,P_+]&=& P_-,\nonumber\\
\relax [P_+,P_-]&=& -\frac{\lambda}{2}C.
\eea
$D$ is the grading operator, the grading $z$ being defined by the commutator
\bea
[D,Z]&=& zZ,
\eea for some given generator $Z$. \par
$H,D,K$ close an $sl(2)$ algebra with $D$ as the Cartan element.\par
$P_\pm$, together with the central charge $C$, realize the $h(1)$ Lie-Heisenberg subalgebra. Since $P_\pm$ have half-integer grading, it is convenient in the following to introduce the notation
\bea\label{equivalPP}
P_\pm\equiv P_{\pm\frac{1}{2}}.
\eea
$P_{\pm\frac{1}{2}}$ induce a closed $osp(1|2)$ algebra, whose even generators are the second-order
differential operators $P_{\pm1}, P_0$, introduced via the anticommutators (for $\delta$ and $\epsilon$ taking values $\pm 1$)
\bea\label{posp}
P_{(\delta+\epsilon)\frac{1}{2}}&=& \{ P_{\delta\frac{1}{2}},P_{\epsilon\frac{1}{2}}\},
\eea
while $P_{\pm\frac{1}{2}}$ belong to the odd sector of $osp(1|2)$. \par
One should note that $[D,P_s]=sP_s$. Since
\bea
[\Omega, P_s] &=& 0,\quad \forall s=\pm1, \pm\frac{1}{2},0,
\eea
the $osp(1|2)$ Lie superalgebra spanned by the $P_s$'s operators is a symmetry superalgebra of the L\'evy-Leblond operator $\Omega$.\par 
A closed symmetry superalgebra $u(1)\oplus sl(2){\supset \hspace{-1em}\hspace{-1pt}+}osp(1|2)$, with $2$ odd and $7$ even generators, is spanned by $H,D,K,C$ and the five $P_s$ operators.\par
We note, as a remark, that if we express the free heat equation
\bea
&\Omega^2\Psi(x,t)=(\lambda\partial_t+\partial_x^2)\Psi(x,t)=0& 
\eea
in the equivalent form
\bea\label{equiv}
-\lambda\partial_t\Psi(x,t)=\partial_x^2\Psi(x,t),
\eea
the operator $P_\frac{1}{2}$ is the square root of the r.h.s. operator in (\ref{equiv}).\par
The L\'evy-Leblond operator $\Omega\equiv{\Omega}_{z(x,t)=1}$ possesses half-integer grading:
\bea\label{halfgrade}
[D,\Omega]&=&\frac{1}{2}\Omega.
\eea
The commutator with $K$ produces an operator with  $-\frac{1}{2}$ grading,
\bea
\relax [K,\Omega] &=& t\Omega\equiv\Omega_{z(x,t)=t},\nonumber\\
\relax [D,t\Omega]&=&-\frac{1}{2}t\Omega.
\eea
It is convenient to set $\Omega\equiv \Omega_\frac{1}{2}$, $t\Omega\equiv \Omega_{-\frac{1}{2}}$.
\par
The commutator
\bea\label{omegatomega}
\relax [\Omega, t\Omega ]&=& \gamma_+\Omega
\eea
gives an operator, $\gamma_+\Omega$, which does not belong to the class of symmetry operators here considered. The reason is that its commutator with $\Omega$ can be expressed as
\bea\label{symomegatomega}
\relax [\Omega,\gamma_+\Omega] &=& (-\lambda\gamma_3+2\gamma_+\partial_x)\Omega\equiv T\Omega,
\eea
where $T= -\lambda\gamma_3+2\gamma_+\partial_x$ is not a matrix-valued function, but a differential operator.\par
On the other hand, $\Omega_{\pm\frac{1}{2}}$ can be taken as the odd sector of an  $osp(1|2)$ superalgebra
whose even generators are the second-order differential operators $\Omega_{\pm 1},\Omega_0$ given by
the anticommutators (for $\delta$ and $\epsilon$ taking values $\pm 1$)
\bea
\Omega_{(\delta+\epsilon)\frac{1}{2}}&=& \{ \Omega_{\delta\frac{1}{2}},\Omega_{\epsilon\frac{1}{2}}\}.
\eea
The non-vanishing commutators of this second $osp(1|2)$ superalgebra are given by
\bea\label{secondosp}
\relax [\Omega_0,\Omega_s]&=& -2s\lambda\Omega_s,\quad\quad s=\pm\frac{1}{2},\pm 1, 0,\nonumber\\
\relax [\Omega_1,\Omega_{-1}]&=& 4\lambda\Omega_0,\nonumber\\
\relax [\Omega_{\pm\frac{1}{2}},\Omega_{\mp 1}]&=& \pm 2\lambda\Omega_{\mp\frac{1}{2}}.
\eea
It follows, as a corollary, that the second-order differential operators $\Omega_{\pm 1},\Omega_0$ 
are symmetry operators (since, e.g., $[\Omega_{-1},\Omega_{\frac{1}{2}}]=-2\lambda t\Omega_{\frac{1}{2}} $).\par
It is worth pointing out that in the Schr\"odinger case, when $\lambda$ is replaced by $\beta J$, see the discussion before equation (\ref{omegabar}), the five symmetry operators closing the $osp(1|2)$ superalgebra are $\Omega_{+\frac{1}{2}}, J\cdot\Omega_{-\frac{1}{2}}, \Omega_{\pm1}, J\cdot\Omega_0$.\par
The action of the $sch(1)$ Schr\"odinger generators $H, D, K, C, P_{\pm \frac{1}{2}}$ on the five operators $\Omega_s$ produces the semidirect superalgebra $sch(1){\supset \hspace{-1em}\hspace{-1pt}+}osp(1|2)$ since the only non-vanishing commutators involving the $\Omega_s$ operators and the Schr\"odinger generators are
\bea
\relax &&[D, \Omega_s]=s\Omega_s,\quad\quad s=\pm \frac{1}{2},\pm 1, 0,\nonumber\\
\relax && [H, \Omega_{-\frac{1}{2}}]=\Omega_{\frac{1}{2}},\quad [H,\Omega_{-1}]= 2\Omega_0,\quad [H,\Omega_0]=\Omega_1,\nonumber\\
\relax && [K, \Omega_{\frac{1}{2}}]=-\Omega_{-\frac{1}{2}},\quad [K,\Omega_{0}]= \Omega_{-1},\quad [K,\Omega_1]=2 \Omega_0.
\eea
Due to the fact that, for any $ s, s'=\pm\frac{1}{2},\pm1,0$,
\bea
\relax [P_s,\Omega_{s'}]&=& 0,
\eea
a first ${\mathbb Z}_2\times {\mathbb Z}_2$-graded Lie superalgebra, whose vector space is spanned by the
$P_s$ and $\Omega_{s'}$ symmetry operators, is obtained. This ${\mathbb Z}_2\times {\mathbb Z}_2$-graded Lie superalgebra is decomposed (see Appendix {\bf A}, equations (\ref{z2z2decomp},\ref{brackets},\ref{z2superalg})) according to
\bea\label{firstz2z2}
&{\cal G}_{00} =\{P_{\pm 1}, P_0, \Omega_{\pm 1}, \Omega_0\}, \quad
{\cal G}_{01} = \{P_{\pm\frac{1}{2}}\},\quad
{\cal G}_{10} =\{ \Omega_{\pm\frac{1}{2}}\},\quad
{\cal G}_{11}= \{\emptyset\},&
\eea
with an empty ${\cal G}_{11}$ sector.\par
A super Schr\"odinger symmetry algebra is obtained. In order to recover it, we need to introduce the symmetry operators arising from the $\Lambda$-sector defined by the equation (\ref{lambda}). The presence of a super Schr\"odinger algebra requires the existence of symmetry operators which are square roots, up to a sign, of $H,K$ entering (\ref{sigmasym},\ref{schr}). In the $4\times 4$ representation, due to the existence of the complex structure $J$, the sign can be flipped (if $Q$ is an operator such that $Q^2=\pm H$, then $J\cdot Q$ is a symmetry operator such that $(J\cdot Q)^2=\mp H$). Our analysis proves the existence of a super Schr\"odinger algebra with at most an ${\cal N}=1$ supersymmetric extension, the only supersymmetric roots being the symmetry operators $Q_\pm$ given by
\bea\label{qplus}
Q_+ &=& \frac{1}{\sqrt{\lambda}}\Lambda_2= \frac{1}{\sqrt{\lambda}}( \gamma_+\partial_t-\lambda\gamma_-) ,\quad \quad Q_+^2= -H,
\eea
and
\bea\label{qminus}
Q_- &=& \frac{1}{\sqrt{\lambda}}(\Lambda_3+{\widetilde \Lambda}_x)=
\frac{1}{\sqrt{\lambda}}(\gamma_+(t\partial_t+x\partial_x+\frac{1}{2})-\gamma_-\lambda t-\gamma_3\frac{\lambda x}{2}),\quad\quad Q_-^2=K.
\eea
Since $Q_\pm$ have half-integer grading,
\bea\label{dosp}
[D,Q_\pm ]&=& \pm \frac{1}{2}Q_\pm,
\eea
as before it is convenient to set $Q_\pm\equiv Q_{\pm\frac{1}{2}}$. \par
The operators $Q_{\pm\frac{1}{2}}$ induce a third $osp(1|2)$ symmetry superalgebra. Their anticommutators
$Q_{(\delta+\epsilon)\frac{1}{2}}=\{ Q_{\delta\frac{1}{2}},Q_{\epsilon\frac{1}{2}}\}$, for $\delta,\epsilon=\pm1$, produce the even generator $Q_{\pm 1}, Q_0$ which coincides, up to normalization, with the $sl(2)$ generators $H,K,D$ given in (\ref{sigmasym}):
\bea\label{qosp}
&
\{Q_{+\frac{1}{2}},Q_{+\frac{1}{2}}\}= -2H,\quad
\{Q_{-\frac{1}{2}},Q_{-\frac{1}{2}}\}= 2K,\quad
\{Q_{+\frac{1}{2}},Q_{-\frac{1}{2}}\} = 2D.&
\eea
The remaining non-vanishing commutators of the third
$osp(1|2)$ superalgebra are
\bea\label{hkosp}
\relax [H, Q_{-\frac{1}{2}}] = Q_{+\frac{1}{2}} &,& [K, Q_{+\frac{1}{2}}]= Q_{-\frac{1}{2}}.
\eea
In the super Schr\"odinger algebra, the two extra operators $P_{\pm\frac{1}{2}}$ enter the even sector and a further symmetry operator $X$, obtained from the commutators
\bea\label{pqx}
[P_{\pm\frac{1}{2}},Q_{\mp\frac{1}{2}}]&=&\pm X, 
\eea
enters the odd sector. We have
\bea
X&=& \frac{1}{\sqrt{\lambda}}{\widetilde \Lambda}_{w(x,t)=1}+\frac{\sqrt\lambda}{2}\Lambda_1= \frac{1}{\sqrt\lambda}(\gamma_+\partial_x-\gamma_3\frac{\lambda}{2}).
\eea
The super Schr\"odinger algebra $ssch(1)$ is decomposed into even and odd sector as
$$
ssch(1)= {\cal G}_0\oplus{\cal G}_1,
$$
with ${\cal G}_0 =\{ H,D,K,C, P_{\pm \frac{1}{2}}\} $ and ${\cal G}_1=\{Q_{\pm \frac{1}{2}},X\}$.\par
The extra non-vanishing (anti)commutators involving the $X$ generator are
\bea
\{X,Q_{\pm\frac{1}{2}}\}&=& - P_{\pm \frac{1}{2}},\nonumber\\
\{X,X\}&=&\frac{\lambda}{2}C.
\eea
It is worth pointing out that the operator $Q_{+\frac{1}{2}}$ is the square root of the operator in the l.h.s. of
the equation (\ref{equiv}). \par
We recovered three independent $osp(1|2)$ symmetry superalgebras induced, respectively, by the three pairs of half-graded odd generators $P_{\pm\frac{1}{2}},\Omega_{\pm\frac{1}{2}},Q_{\pm\frac{1}{2}}$. We already discussed the compatibility of the $osp(1|2)$ superalgebras obtained from the two pairs $P_{\pm\frac{1}{2}},\Omega_{\pm\frac{1}{2}}$. The last step consists in discussing the compatibility of the $osp(1|2)$ superalgebras induced by the two pairs $ Q_{\pm\frac{1}{2}},\Omega_{\pm\frac{1}{2}} $ and $Q_{\pm\frac{1}{2}},P_{\pm\frac{1}{2}}$. \par
It is easily shown that an algebra which presents both pairs of  $ Q_{\pm\frac{1}{2}},\Omega_{\pm\frac{1}{2}} $ generators brings us beyond the scheme of symmetry algebras discussed in the present paper. Indeed, their
repeated (anti)commutators produce higher derivative operators which do not satisfy equations  (\ref{sigma}) and (\ref{lambda}) with matrix-valued functions on their right hand sides.\par
On the other hand, requiring the presence of the two pairs of operators $ Q_{\pm\frac{1}{2}}, P_{\pm\frac{1}{2}} $, a ${\mathbb Z}_2\times{\mathbb Z}_2$-graded Lie superalgebra 
(following the definition in Appendix {\bf A}) of symmetry operators is naturally induced. This superalgebra, ${\cal G}_{{\mathbb Z}_2\times{\mathbb Z}_2}$,  is decomposed according to ${\cal G}_{{\mathbb Z}_2\times{\mathbb Z}_2}={\cal G}_{00}\oplus {\cal G}_{01}\oplus {\cal G}_{10}\oplus {\cal G}_{11}$, with their respective sectors spanned by the generators
\bea\label{z2z2sym}
{\cal G}_{00} &=& \{ H, D, K, P_{\pm 1}, P_0\},\nonumber\\
{\cal G}_{01} &=& \{ P_{\pm \frac{1}{2}}\},\nonumber\\
{\cal G}_{10} &=& \{ Q_{\pm{\frac{1}{2}}}, X_{\pm\frac{1}{2}}\},\nonumber\\
{\cal G}_{11} &=& \{ X\}.
\eea
The extra generators we have yet to introduce are the second-order differential operators $X_{\pm\frac{1}{2}}$, obtained from
the anticommutators
\bea
X_{\pm\frac{1}{2}}&=&\{X,P_{\pm\frac{1}{2}}\}.
\eea
Some comments are in order. The ${\mathbb Z}_2\times{\mathbb Z}_2$-graded Lie superalgebra (\ref{z2z2sym}) is closed under (anti)-commutators and, furthermore, the ${\mathbb Z}_2\times{\mathbb Z}_2$-graded Jacobi identities are fulfilled, as explicitly verified.\par
It is important to point out that
 (\ref{z2z2sym})  is a ${\mathbb Z}_2\times{\mathbb Z}_2$-graded Lie superalgebra of symmetry operators
(as defined in Section {\bf 3}). The last check consists in verifying that the anticommutators of $X_{\pm\frac{1}{2}}$ with $\Omega\equiv\Omega_{+\frac{1}{2}}$ are vanishing:
\bea
&\{\Omega,X\}=\{\Omega,X_{\pm\frac{1}{2}}\}=0.&
\eea
For completeness we present the list of the remaining (besides the ones already introduced) non-vanishing (anti)commutators involving the
(\ref{z2z2sym}) generators. We have
\medskip\noindent
\begin{equation}
  \begin{array}{lcl}
    [D, P_s] = s P_s, & & [D, X_{\pm\frac{1}{2}}] = \pm\frac{1}{2} X_{\pm\frac{1}{2}}, \\[5pt]
    [H, P_m] = (1-m) P_{m+1}, & & 
    [H, X_{-\frac{1}{2}}] = X_{\frac{1}{2}}, \\[5pt]
    [K, P_m] = (1+m) P_{m-1}, & & 
    [K, X_{\frac{1}{2}}] = X_{-\frac{1}{2}}, \\[5pt]
    [P_1, P_{-1}] = -4 \lambda P_0, & & [P_{\pm 1}, P_{\mp \frac{1}{2}}] = \mp 2\lambda P_{\pm\frac{1}{2}}, \\[5pt]
    [P_{\pm 1}, Q_{\mp \frac{1}{2}}] = \pm 2 X_{\pm \frac{1}{2}}, & & 
    [P_{\pm 1}, X_{\mp \frac{1}{2}}] = \mp 2\lambda X_{\pm \frac{1}{2}}, \\[5pt]
    [P_0, P_s] = 2s \lambda P_s, & & [P_0, Q_{\pm \frac{1}{2}} ] = \mp X_{\pm\frac{1}{2}}, \\[5pt]
    [P_0, X_{\pm\frac{1}{2}}] = \pm \lambda X_{\pm\frac{1}{2}}, & & 
    [P_{\pm\frac{1}{2}}, Q_{\mp\frac{1}{2}} ] = \pm X, \\[5pt]
    [P_{\pm\frac{1}{2}}, X_{\mp\frac{1}{2}}] = \mp \lambda X, & &
    \{ X_{\pm\frac{1}{2}}, X_{\pm\frac{1}{2}} \} = \lambda P_{\pm 1}, \\[5pt]
    \{ X_{\frac{1}{2}}, X_{-\frac{1}{2}} \} = \lambda P_0, & & 
    \{ Q_{\pm\frac{1}{2}}, X_{\pm\frac{1}{2}} \} = -P_{\pm 1}, \\[5pt]
    \{ Q_{\pm\frac{1}{2}}, X_{\mp\frac{1}{2}} \} = - P_0, & & 
    \{ X, X_{\pm\frac{1}{2}} \} = \lambda P_{\pm\frac{1}{2}}.
  \end{array}
\end{equation}
where $ m = \pm 1, 0 $ and $ s=\pm\frac{1}{2},\pm 1, 0$.\par
It is worth pointing out that adding the identity operator $C$ to the ${\cal G}_{00}$ sector, we obtain the ${\mathbb Z}_2\times {\mathbb Z}_2$-graded Lie superalgebra $u(1)\oplus {\cal G}_{{\mathbb Z}_2\times{\mathbb Z}_2}$. The ${\cal N}=1$ super Schr\"odinger algebra $ssch(1)$ is spanned by a subset of the $u(1)\oplus {\cal G}_{{\mathbb Z}_2\times{\mathbb Z}_2}$ generators. Even so, $ssch(1)$ is not a
$u(1)\oplus {\cal G}_{{\mathbb Z}_2\times{\mathbb Z}_2}$ subalgebra (since for $ssch(1)$ the brackets involving, e.g.,
$P_{\pm\frac{1}{2}}$ are defined with commutators).

\section{Symmetries of the L\'evy-Leblond square root of the heat equation with quadratic potential}

The Schr\"odinger algebra is the maximal kinetic symmetry algebra of the heat or of the Schr\"odinger equation, see \cite{nie,boy,top}, for three type of potentials, constant, linear or quadratic. The last case corresponds to the harmonic oscillator. We investigate here the symmetries of the the associated L\'evy-Leblond equation for the heat equation with quadratic potential. On the basis of the construction of Section {\bf 2}, the minimal matrix differential operator is expressed as a $4\times4$ matrix. The quadratic potential is recovered from the linear function $f(x)=\omega x$ entering (\ref{heat}). Without loss of generality we can set $\lambda=\omega=1$ and take the block-antidiagonal L\'evy-Leblond operator ${\widetilde\Omega}$ to be given by
\bea\label{llpot}
{\widetilde\Omega}&=& ( e_{14}-e_{32})\partial_t+(e_{13}-e_{24}-e_{31}+e_{42})\partial_x+ x(e_{13}+e_{24}+e_{31}+e_{42})-e_{23}+e_{41},
\eea
where $e_{ij}$ denotes the matrix with entry $1$ at the cross of the $i$-th row and $j$-th column and $0$ otherwise.\par
In this basis its squared operator ${\widetilde\Omega}^2$ is
\bea
{\widetilde\Omega}^2&=& (e_{11}+e_{22}+e_{33}+e_{44})(\partial_t-\partial_x^2+x^2) +e_{11}+e_{44}-e_{22}-e_{33}.
\eea
The exhaustive list of first-order differential symmetry operators $\Sigma$'s satisfying
\bea
[\Sigma,{\widetilde\Omega}]&=&\Phi_\Sigma(x,t)\cdot {\widetilde\Omega},
\eea
for some $4\times4$ matrix-valued functions $\Phi_\Sigma(x,t)$ in the $x,t$ coordinates, can be computed with lengthy but straightforward methods. We present here the complete list of symmetry operators (for simplicity we leave as an exercise for the Reader the computation of their associated matrix-valued functions $\Phi_\Sigma(x,t)$'s).\par
The symmetry operators can be split into the two big classes of block-diagonal and block-antidiagonal operators. In both cases we have $12$ fixed symmetry operators (up to normalization) plus two extra sets of operators depending on an unconstrained function (denoted as $z(x,t)$) of the $x,t$ coordinates.\par
The $12$ fixed block-diagonal symmetry operators can be conveniently presented as
\bea\label{sympot1}
\Sigma_1&=& e^{4t}((e_{22}+e_{33}-2xe_{34})\partial_t+(e_{21}-e_{43}+2x(e_{22}+e_{44}))\partial_x+\nonumber\\&&xe_{21}+2x^2e_{22}+4e_{33}-8xe_{34}-xe_{43}+2(x^2+1)e_{44}),\nonumber\\
\Sigma_2&=& e^{4t}((e_{11}+e_{44}+2xe_{34})\partial_t+(e_{43}-e_{21}+2x(e_{11}+e_{33}))\partial_x+\nonumber\\
&&(2+2x^2)e_{11}-xe_{21}+2x^2e_{33}+xe_{43}),\nonumber\\
\Sigma_3&=& e^{-4t} ( (e_{22}+e_{33}+2xe_{34})\partial_t+(e_{21}-e_{43}-2x(e_{22}+e_{44}))\partial_x +\nonumber\\&& -3x e_{21}-2x^2e_{22}+(4x^2-2)e_{33}-xe_{43}+2x^2e_{44}),\nonumber\\
\Sigma_4&=& e^{-4t}((e_{11}+e_{44}-2xe_{34})\partial_t+(e_{34}-e_{21}-2x(e_{11}+e_{33}))\partial_x +\nonumber\\
&&(2x^2-4)e_{11}-8xe_{12}+3xe_{21}+(4x^2-2)e_{22}-2x^2e_{33}+xe_{43}),\nonumber\\
\Sigma_5&=& e^{2t}((e_{11}+e_{22}+e_{33}+e_{44})(\partial_x+x) -2e_{34}),\nonumber\\
\Sigma_6&=& e^{-2t}((e_{11}+e_{22}+e_{33}+e_{44})(\partial_x-x) +2e_{12}),\nonumber\\
\Sigma_7&=&(e_{11}+e_{44})\partial_t+(e_{43}-e_{21})\partial_x+x(e_{21}+e_{43}),\nonumber\\
\Sigma_8&=& (e_{22}+e_{33})\partial_t +(e_{21}-e_{43})\partial_x-x(e_{21}+e_{43}),\nonumber\\
\Sigma_9&=& e_{11}+e_{22},\nonumber\\
\Sigma_{10}&=&e_{33}+e_{44},\nonumber\\
\Sigma_{11}&=&e^{2t}(  e_{34}\partial_t-(e_{22}+e_{44})\partial_x-e_{21}+2e_{34}-x(e_{22}+e_{44})     ),\nonumber\\
\Sigma_{12}&=&e^{-2t}( e_{34}\partial_t-(e_{22}+e_{44})\partial_x-e_{21}+x(e_{44}+2e_{33}-e_{22})      ).
\eea
The $12$ fixed block-antidiagonal symmetry operators can be conveniently presented as
\bea\label{sympot2}
\Sigma_{13}&=& e^{6t}((e_{13}-e_{41}-4xe_{32})\partial_t+(4x(e_{13}-e_{31})-(e_{23}+e_{41}))\partial_x+\nonumber\\&&(8x^2+4)e_{13}-
(12x+8x^3)e_{14}-3xe_{23}+(4x^2+2)e_{24}+4x^2e_{31}+xe_{41}),\nonumber\\
\Sigma_{14}&=& e^{-6t}( (e_{24}-e_{31}-4xe_{32})\partial_t+(e_{23}+e_{41}+4x(e_{42}-e_{24}))\partial_x+\nonumber\\
&&-3xe_{23}+4x^2e_{24}+4(1-x^2)e_{31}+(12x-8x^3)e_{32}+xe_{41}+2e_{42}),\nonumber\\
\Sigma_{15}&=&e^{4t}(-e_{32}\partial_t+(e_{13}-e_{31})\partial_x+3xe_{13}-(4x^2+2)e_{14}-e_{23}+2xe_{24}+xe_{31}),\nonumber\\
\Sigma_{16}&=&e^{-4t}(e_{32}\partial_t+(e_{24}-e_{42})\partial_x+e_{23}-xe_{24}+2xe_{31}+(4x^2-2)e_{32}+xe_{42}       ),\nonumber\\
\Sigma_{17}&=& e^{2t}\left((e_{24}-e_{31})\partial_t+e_{23}\partial_x+x(e_{23}+e_{41})-2e_{31}\right),\nonumber\\
\Sigma_{18}&=& e^{2t}((e_{13}-e_{42}-2xe_{32})\partial_t-(e_{23}+e_{41}+2x(e_{31}-e_{13}))\partial_x+x(e_{41}-e_{23})+2x^2(e_{13}+e_{31})),\nonumber\\
\Sigma_{19}&=&e^{2t}(  e_{13}+e_{24}-2xe_{14}   ),\nonumber\\ 
\Sigma_{20}&=& e^{-2t}((e_{13}-e_{42})\partial_t-(e_{23}+e_{41})\partial_x-2e_{13}+x(e_{23}+e_{41})),\nonumber\\
\Sigma_{21}&=& e^{-2t}\left((e_{24}-e_{31}-2xe_{32})\partial_t+(-2x(e_{24}-e_{42})+e_{23}+e_{41})\partial_x
+2x^2(e_{24}+e_{42})+x(e_{41}-e_{23})\right),\nonumber\\
\Sigma_{22}&=&e^{-2t}(e_{31}+e_{42}+2xe_{32}       ),\nonumber\\ 
\Sigma_{23}&=& (e_{13}+e_{24}-e_{31}-e_{42})\partial_x +(e_{13}-e_{24}+e_{31}-e_{42})x,\nonumber\\
\Sigma_{24}&=&e_{32}\partial_t+(e_{24}-e_{42})\partial_x+e_{23}-x(e_{24}+e_{42}).
\eea
The $4$ (two block-diagonal and two block-antidiagonal) extra sets of symmetry operators depending on the unconstrained functions $z_i(x,t)$, $i=1,2,3,4$, are
\bea\label{sympot3}
{\widetilde\Sigma}_{1, z_1(x,t)}&=& z_1(x,t)((e_{12}+e_{34})\partial_t +(e_{11}-e_{22}+e_{33}-e_{44})\partial_x-(e_{21}+e_{43})-x(e_{11}+e_{22}-e_{33}-e_{44})),\nonumber\\
{\widetilde\Sigma}_{2,z_2(x,t)}&=&z_2(x,t)((e_{12}-e_{34})\partial_x+e_{11}-e_{33}+x(e_{12}+e_{34})),\nonumber\\
{\widetilde\Sigma}_{3,z_3(x,t)}&=& z_3(x,t)((e_{14}-e_{32})\partial_t+(e_{13}-e_{24}-e_{31}+e_{42})\partial_x-e_{23}+e_{41}+x(e_{13}+e_{24}+e_{31}+e_{42})),\nonumber\\
{\widetilde \Sigma}_{4,z_4(x,t)}&=& z_4(x,t)((e_{14}+e_{32})\partial_x+e_{13}+e_{31}+x(e_{32}-e_{14})).
\eea
The L\'evy-Leblond operator ${\widetilde\Omega}$ is recovered from ${\widetilde\Sigma}_{3,z_3(x,t)}$ by setting $z_3(x,t)=1$. We have, indeed,
\bea
{\widetilde\Omega}&=& {\widetilde\Sigma}_{3,z_3(x,t)=1}.
\eea
Some of the block-diagonal symmetry operators have special meaning: the identity ${\mathbb I}_4$ is given by the combination ${\mathbb I}_4=\Sigma_{9}+\Sigma_{10}$, while the fermion-number symmetry operator $N_f$ is the difference $N_f=\Sigma_9-\Sigma_{10}$. The time-derivative is a symmetry operator and,  as it is the case for the Schr\"odinger's equation of the  harmonic oscillator, it can be used to define a grading operator $D$.  We can set
\bea
D&=& \frac{1}{4}(\Sigma_7+\Sigma_8)= {\mathbb I}_4\cdot\frac{1}{4}\partial_t.
\eea
A natural Schr\"odinger symmetry algebra is encountered. The identification (up to normalization) of its symmetry generators at degree $\pm1, \pm\frac{1}{2},0$ is given by
\bea\label{schpot}
&+1: \Sigma_1+\Sigma_2\equiv H,\quad +\frac{1}{2}: \Sigma_5\equiv P_{+\frac{1}{2}},\quad 0: D, C\equiv{\mathbb I}_4, \quad-\frac{1}{2}: \Sigma_6\equiv P_{-\frac{1}{2}},\quad -1: \Sigma_3+\Sigma_4\equiv K.& \nonumber\\&&
\eea
$\Sigma_5\equiv P_{\frac{1}{2}},\Sigma_6\equiv P_{-\frac{1}{2}}$ are, respectively, the creation/annihilation operators, while $H,D,K$ are the $sl(2)$ subalgebra generators.\par
A first difference with respect to the symmetry algebra of the free L\'evy-Leblond equation is already encountered
at this level. In the free case the L\'evy-Leblond operator possesses half-integer grading ($=\frac{1}{2}$) with respect to the $sl(2)$ Cartan generator of the Schr\"odinger subalgebra. In the present case ${\widetilde\Omega}$ has zero grading
\bea\label{0grade}
[D,{\widetilde\Omega}]&=&0.
\eea
This implies, as a a corollary, that for the quadratic potential there is no $osp(1|2)$ symmetry superalgebra induced by $\Omega_{\pm\frac{1}{2}}$.\par
The extra question to be investigated is whether, in the presence of a non-trivial potential, the L\'evy-Leblond operator admits a super-Schr\"odinger invariance.\par
In principle one can repeat the same steps as in the free case and compute the most general solution obtained from the anti-commutators $\{\Lambda,{\widetilde\Omega}\}=\Phi_\Lambda(x,t)\cdot{\widetilde\Omega}$.
On the other hand, since these computations are rather cumbersome, it is preferable to address the question with a different procedure, based on the key observation that the existence of the super-Schr\"odinger symmetry requires the presence of at least one operator which is the square root (up to normalization) of $H=  \Sigma_1+\Sigma_2$. Requiring $Q^2=kH$ for some real $k\neq 0$ produces the most general solution
\bea\label{susch}
Q&=& e^{2t}\left(\begin{array}{cccc}0&\partial_t+2x\partial_x+2x^2&r_3&-2r_3x\\
r_1r_2&0&0&-r_1r_3\\
0&0&2r_2x&r_1\partial_t+2r_1x\partial_x+(2r_1-4r_2)x^2+2r_1\\0&0&r_2&-2r_2x
\end{array}\right),
\eea
depending on three real parameters $r_i$  ($i=1,2,3$) and with the identification $k=r_1r_2$. Therefore, both
$r_1,r_2$ need to be non-vanishing. Once identified the most general operator $Q$, we need to verify whether it belongs to the class of L\'evy-Leblond symmetry operators. This means that a matrix-valued function $\Phi(x,t)$ should be found for
\bea\relax
either \quad [Q,{\widetilde\Omega}]=\Phi(x,t)\cdot{\widetilde\Omega} & {\textstyle or} & \{Q,{\widetilde\Omega}\}=\Phi(x,t)\cdot{\widetilde\Omega}.
\eea
It is easily checked that, for $r_{1,2}\neq 0$, no such matrix-valued function exists, implying that the $r_{1,2}\neq 0$ operators in (\ref{susch}) are not symmetry generators of the L\'evy-Leblond operator (\ref{llpot}) with non-trivial potential.\par
In the presence of a non-trivial potential there is no analog of the $osp(1|2)$ symmetry generated by
the  $Q_{\pm\frac{1}{2}}$ symmetry operators of the free case.\par
In the free case we discussed three separated $osp(1|2)$ symmetry superalgebras, induced by
$P_{\pm\frac{1}{2}},Q_{\pm\frac{1}{2}},\Omega_{\pm\frac{1}{2}}$, respectively. In the presence of the non-trivial potential we obtain a single $osp(1|2)$ symmetry from the (anti)commutators of the creation/annihilation operators $P_{\pm\frac{1}{2}}$ entering (\ref{schpot}), with the even sector being given by
$P_{\pm 1}=2P_{\pm\frac{1}{2}}^2$, $P_0=\{ P_{+\frac{1}{2}} ,P_{-\frac{1}{2}}\}$. All these operators commute with ${\widetilde\Omega}$:
\bea
[{\widetilde\Omega}, P_s]&=&0,\quad\quad s=0,\pm\frac{1}{2},\pm 1.
\eea
Together with the extra generator $H,D,K, C$ in (\ref{schpot}), we therefore find a symmetry superalgebra for ${\widetilde\Omega}$, given by $u(1)\oplus sl(2){\supset \hspace{-1em}\hspace{-1pt}+}osp(1|2)$. It contains two odd generators $P_{\pm\frac{1}{2}}$, while the $7$ remaining generators span the even sector.
\par
The investigation and identification of other closed finite symmetry (super)algebras recovered from the (\ref{sympot1},\ref{sympot2},\ref{sympot3}) symmetry operators is beyond the scope of the present paper.

\section{Conclusions}

We highlight the main  results of the paper. \par We proved, for the L\'evy-Leblond square root of the free heat or the free Schr\"odinger equation, the existence of a symmetry of first-order and second-order differential operators closing a finite ${\mathbb Z}_2\times {\mathbb Z}_2$-graded Lie superalgebra. Let's take, for the sake of clarity, the $1+1$-dimensional Schr\"odinger equation. It can be equivalently written in two forms, either
\bea\label{schrone}
(i\partial_t+\partial_x^2)\Psi(x,t)&=&0,
\eea
or
\bea\label{schrtwo}
i\partial_t\Psi(x,t)&=& -\partial_x^2\Psi(x,t),
\eea
where $2\times 2$ diagonal operators (the identity ${\mathbb I}_2$ is dropped for simplicity) act upon the
$2$-column complex vector $\Psi(x,t)$. \par
Three independent sets of $osp(1|2)$ symmetry superalgebras are encountered. They are generated by the first-order,  odd, symmetry differential operators $\Omega_{\pm\frac{1}{2}}$, $P_{\pm\frac{1}{2}}$, $Q_{\pm\frac{1}{2}}$, respectively.\par
$\Omega_{+\frac{1}{2}}$ is the L\'evy-Leblond operator we initially started with; it is the square root of the operator entering
(\ref{schrone}). $P_{+\frac{1}{2}}$ ($Q_{+\frac{1}{2}}$) is the square root of the operator entering the right hand side (respectively, the left hand side)  of equation (\ref{schrtwo}). We investigated the mutual compatibility conditions for these $osp(1|2)$ superalgebras. Since the $\Omega_{\pm\frac{1}{2}}$ operators commute with the $P_{\pm\frac{1}{2}}$ operators, 
a first ${\mathbb Z}_2\times {\mathbb Z}_2$-graded Lie superalgebra, given by (\ref{firstz2z2}), is encountered.  Its ${\cal G}_{11}$ sector is empty, while its vector space is spanned by the
$P_s$ and $\Omega_{s'}$ symmetry operators ($s,s'=\pm\frac{1}{2},\pm1,0$).\par 
 The $\Omega_{\pm \frac{1}{2}}$ operators, together with the $Q_{\pm\frac{1}{2}}$ operators, produce an algebra containing higher-order (degree $\geq 3$) differential operators. \par
A finite, closed, non-trivial ${\mathbb Z}_2\times {\mathbb Z}_2$-graded Lie superalgebra (spanned by the $13$ generators recovered from (\ref{z2z2sym})) is discovered by requiring the presence of the two pairs of operators, $P_{\pm\frac{1}{2}}$ and $Q_{\pm\frac{1}{2}}$. \par The  (\ref{schrone},\ref{schrtwo}) system also possesses a (maximally ${\cal N}=1$) ${\mathbb Z}_2$-graded super Schr\"odinger algebra symmetry.
\par
We completed the ($1+1$)-dimensional investigation by looking at the symmetries of the L\'evy-Leblond square root of the ($1+1$)-dimensional heat equation with quadratic potential. This system, defined by the L\'evy-Leblond operator ${\widetilde \Omega}$, possesses a symmetry closing the Schr\"odinger algebra $sch(1)$. We proved, on the other hand, by exhaustive computations, that the non-vanishing potential does not allow
a graded extension (neither ${\mathbb Z}_2$- nor ${\mathbb Z}_2\times{\mathbb Z}_2$-) of the Schr\"odinger algebra. The main difference, with respect to the free case, is due to the fact that ${\widetilde \Omega}$ commutes with the Cartan generator $D$ of the $sl(2)\subset sch(1)$ subalgebra, see (\ref{0grade}). Measured by $D$, ${\widetilde \Omega}$ is a $0$-grade operator.  In the free case, the
L\'evy-Leblond operator ${\Omega_{+\frac{1}{2}}}$ possesses the half-integer $+\frac{1}{2}$-grade when measured by the corresponding $D$ operator, see (\ref{halfgrade}).
\par 
Some comments are in order. 
To our knowledge, this is the first time that a ${\mathbb Z}_2\times {\mathbb Z}_2$-graded Lie (super)algebra is encountered in the context of symmetries of partial differential equations. This feature could  prove to be significant for the community of physicists (searching for applications of) and mathematicians (developing the mathematical structure of) graded color Lie (super)algebras. It opens the possibility,
e.g., to use them as spectrum-generating algebras for suitable dynamical systems.\par
The constructions of (both ${\mathbb Z}_2$- and ${\mathbb Z}_2\times{\mathbb Z}_2$-) graded extensions of Lie symmetries require two types of symmetry operators, the ones obtained from the commutator condition (\ref{comm}) and those obtained from the anticommutator condition (\ref{anticomm}).\par
In Appendix {\bf B} we constructed a ${\mathbb Z}_2\times{\mathbb Z}_2$-graded Lie superalgebra symmetry of the L\'evy-Leblond operator associated with the free heat equation in ($1+2$)-dimensions. 
A new feature appears. The ${\mathbb Z}_2\times {\mathbb Z}_2$-graded symmetry allows to explain the existence of first-order differential symmetry operators which do not belong to the two-dimensional super Schr\"odinger algebra.  \par
It is easily realized that our results have a more general validity, so that (\ref{schrone},\ref{schrtwo}) and (\ref{d2levyl}) are the prototypes of a class of ${\mathbb Z}_2\times{\mathbb Z}_2$-graded invariant equations.  The formal demonstration, based on the systematic Clifford algebra tools discussed in section {\bf 2}, of the existence of a ${\mathbb Z}_2\times{\mathbb Z}_2$-graded Lie superalgebra symmetry for the L\'evy-Leblond square roots of the free heat or Schr\"odinger equations in ($1+d$)-dimension, for an arbitrary number of space dimensions $d$,  will be presented elsewhere. We are reporting the  ${\mathbb Z}_2\times{\mathbb Z}_2$-graded symmetry
of the original L\'evy-Leblond equation (the square root of the ($1+3$)-dimensional free Schr\"odinger equation) in \cite{aktt}, a paper based on a talk given at the 2016 ICGTMP (Group 31 Colloquium). \par
Finally, as discussed in Appendix {\bf C}, future investigations of ${\mathbb Z}_2\times {\mathbb Z}_2$-graded symmetries of partial differential equations should systematically study the different  possible graded Lie (super)algebras recovered from different assignment of gradings to the symmetry operators. 
 \par
We already anticipated in the Introduction the direction of our investigations which are currently under way,
concerning the possible arising of emergent ${\mathbb Z}_2\times {\mathbb Z}_2$-graded symmetry, beyond the realm of the L\'evy-Leblond equations, for the different classes of theories discussed in references \cite{wit}-\cite{vit2}. An intriguing possibility is that the ${\mathbb Z}_2\times {\mathbb Z}_2$-graded symmetry, employed as a spectrum-generating algebra, could allow derive the full spectrum of a theory from a single irreducible lowest weight representation, while the superalgebra (contained as subalgebra inside the ${\mathbb Z}_2\times {\mathbb Z}_2$-structure) only possesses partial information, presenting the spectrum as the decomposition into a direct sum of (several) of its irreducible representations.
{}~\par{}~\par 
\textcolor{black}{}
\par {\Large{\bf Acknowledgments}}
{}~\par{}~\par 
\textcolor{black}{}
Z.K. and F.T. are grateful to the Osaka Prefecture University, where this work was initiated, for hospitality.
F.T. received support from CNPq (PQ Grant No. 306333/2013-9). 
N. A. is supported by the  grants-in-aid from JSPS (Contract No. 26400209).
\\ {~}~{~}
\\ {~}~{~}
\renewcommand{\theequation}{A.\arabic{equation}}
\setcounter{equation}{0}
 
{\Large{\bf Appendix A: On ${\mathbb Z}_2\times{\mathbb Z}_2$-graded color Lie (super)algebras and their relations with quaternions and split-quaternions}}\par
~\par
To make the paper self-contained we collect here the main properties of the ${\mathbb Z}_2\times{\mathbb Z}_2$-graded color Lie algebras and superalgebras induced by a composition law; we make also explicit their relations with both the division algebra of quaternions and its split version.\par

A ${\mathbb Z}_2\times {\mathbb Z}_2$-graded color(super)algebra ${\cal G}$ is decomposed according to 
\bea \label{z2z2decomp}
{\cal G}&=&{\cal G}_{00}\oplus{\cal G}_{01}\oplus {\cal G}_{10}\oplus {\cal G}_{11}
.
\eea
For the generators $x_{\vec \alpha}, x_{\vec \beta}\in {\cal G}$, with ${\vec \alpha}, \vec{\beta}$ labeling the grading, the ${\cal G}\times {\cal G}\rightarrow {\cal G}$ brackets are
\bea\label{brackets}
(x_{\vec \alpha}, x_{\vec \beta}) &=& x_{\vec \alpha} x_{\vec \beta} - (-1)^{({\vec\alpha}\cdot{\vec \beta})}x_{\vec\beta}x_{\vec\alpha},
\eea
where $x_{\vec\alpha}x_{\vec\beta}$ is a composition law and, for the ${\mathbb Z}_2\times {\mathbb Z}_2$-graded color algebra, the inner product is defined by
\bea\label{z2alg}
({\vec\alpha}\cdot{\vec \beta})&=& \alpha_1\beta_2-\alpha_2\beta_1,
\eea
while, for the ${\mathbb Z}_2\times {\mathbb Z}_2$-graded color superalgebra, the inner product
is given by
\bea\label{z2superalg}
({\vec\alpha}\cdot{\vec \beta})&=& \alpha_1\beta_1+\alpha_2\beta_2.
\eea
For both ${\mathbb Z}_2\times {\mathbb Z}_2$-graded color algebra and superalgebra, the grading 
$deg(x_{\vec \alpha}, x_{\vec \beta})$ of the $(x_{\vec \alpha}, x_{\vec \beta})$ bracket is
\bea
deg(x_{\vec \alpha}, x_{\vec \beta}) &=& {\vec \alpha }+{\vec \beta}.
\eea
${\mathbb Z}_2\times {\mathbb Z}_2$-graded Jacobi identities are imposed, in both algebra and superalgebra cases, through
\bea\label{gradedjac}
(-1)^{({\vec \alpha} \cdot{\vec \gamma}) } (x_{\vec \alpha},(x_{\vec \beta}, x_{\vec \gamma})) +
(-1)^{({\vec \beta} \cdot{\vec \alpha})} (x_{\vec \beta},(x_{\vec \gamma}, x_{\vec \alpha})) +
(-1)^{({\vec \gamma} \cdot{\vec \beta})} (x_{\vec \gamma},(x_{\vec \alpha}, x_{\vec \beta})) &=&0.
\eea
Some comments are in order. In both algebra and superalgebra cases, the ${\cal G}_{00}$-graded sector of ${\cal G}$  is singled out with respect to the three other sectors. For what concerns the three remaining sectors we have that\\
- for the color algebra, all three sectors  ${\cal G}_{01}$, ${\cal G}_{10}$ and ${\cal G}_{11}$ are on equal footing while,\\
- for the color superalgebra, the ${\cal G}_{11}$ sector is singled out with respect to the  ${\cal G}_{01}$ and ${\cal G}_{10}$ sectors (which are equivalent and can be interchanged).\par
The basic examples of ${\mathbb Z}_2\times {\mathbb Z}_2$-graded Lie algebras and superalgebras are induced by the quaternions and the split-quaternions. We recall here their basic features.\par
The division algebra of the quaternions (over $\mathbb R$) is given by the generators $e_0$ (the identity) and the three imaginary roots $e_i$, $i=1,2,3$,
with composition law
\bea\label{quatcomplaw}
e_i\cdot e_j &=& -\delta_{ij}e_0 +\epsilon_{ijk}e_k,
\eea
for the totally antisymmetric tensor $\epsilon_{123}=\epsilon_{231}=\epsilon_{312}=1$.\par
One should note that the three imaginary quaternions are on equal footing.\par
The split-quaternions (over $\mathbb R$) are given, see \cite{mcC}, by the generators ${\widetilde e}_0$ (the identity) and ${\widetilde e}_i$, $i=1,2,3$, with composition law
\bea\label{splitquatcomplaw}
{\widetilde e}_i\cdot{\widetilde e}_j&=& -\eta_{ij}{\widetilde e}_0+ {\widetilde \epsilon}_{ijk} {\widetilde e}_k.
\eea
The metric $\eta_{ij}$ is diagonal and satisfies
\bea
\eta_{11}=\eta_{22}=-\eta_{33}&=& 1,
\eea
while the totally antisymmetric tensor ${\widetilde \epsilon}_{ijk}$ is 
\bea
{\widetilde \epsilon}_{123}=-{\widetilde \epsilon}_{231}=-{\widetilde\epsilon}_{312} &=& - 1.
\eea
For split-quaternions, only the generators ${\widetilde e}_1$, ${\widetilde e}_2$ are on equal footing and can be interchanged.\par
Different Lie (super)algebras or color Lie (super)algebras  can be defined in terms of both quaternions and split-quaternions; their brackets are given by either commutators or anticommutators obtained from the  composition laws
((\ref{quatcomplaw}) for quaternions and (\ref{splitquatcomplaw}) for split-quaternions). In all cases the graded Jacobi identities are satisfied.\par
Starting from either the quaternions or the split-quaternions, the following algebraic structures can be defined:\\
{\em i}) a Lie-algebra structure with brackets defined by the commutators;\\
{\em ii}) ${\mathbb Z}_2$-graded Lie algebra structures with brackets defined by the appropriate (anti)commutators;\\
{\em iii})  a ${\mathbb Z}_2\times {\mathbb Z}_2$-graded color Lie algebra structure defined by the  (anti)commutators given by (\ref{brackets}) and (\ref{z2alg});\\
{\em iv}) ${\mathbb Z}_2\times {\mathbb Z}_2$-graded  color Lie superalgebras structures defined by the appropriate (anti)commutators given by (\ref{brackets}) and (\ref{z2superalg}) .\\
In all cases the identity (either $e_0$ or ${\widetilde e}_0$) belongs to the even sector
(${\cal G}_{00}$ for the ${\mathbb Z}_2\times{\mathbb Z}_2$ gradings).
\par
An important remark is that in the quaternionic case, 
since all three imaginary roots are on equal footing, there exists a unique
assignment of the graded  (super)algebras in all four cases above. Without loss of generality
we can assign, for the $ii$ case, $e_0,e_3\in{\cal G}_0$ and $e_1,e_2\in{\cal G}_1$; for both $iii$ and $iv$ cases the assignment, without loss of generality, can be assumed to be $e_1\in {\cal G}_{01}, e_2\in {\cal G}_{10}, e_3\in{\cal G}_{11}$.\par
For split-quaternions, since ${\widetilde e}_3$ is singled out with respect to ${\widetilde e}_1, {\widetilde e}_2$,  a unique assignment is only present in the Lie algebra cases {\em i} and {\em iii}. In the supercases
{\em ii} and {\em iv} the assignment of the grading depends on a $\theta$-angle. We have indeed that\\
- for the ${\mathbb Z}_2$ superalgebra case, inequivalent consistent assignments are given by ${\widetilde e}_0, \cos\theta {\widetilde e}_3 +\sin\theta{\widetilde e}_1\in {\cal G}_0$ and ${\widetilde e}_3, -\sin\theta {\widetilde e}_3 +\cos\theta{\widetilde e}_1\in {\cal G}_1$;\\
- for the  ${\mathbb Z}_2\times {\mathbb Z}_2$ color superalgebra case, inequivalent consistent assignments are given by $\cos\theta {\widetilde e}_3 +\sin\theta{\widetilde e}_1\in {\cal G}_{11}$, $-\sin\theta {\widetilde e}_3 +\cos\theta{\widetilde e}_1\in {\cal G}_{01}$, ${\widetilde e}_2\in{\cal G}_{10}$.
%
%
~
\\ {~}~{~}
\\ {~}~{~}

\renewcommand{\theequation}{B.\arabic{equation}}
\setcounter{equation}{0}

{\Large{\bf Appendix B: {${\mathbb Z}_2\times {\mathbb Z}_2$-graded Lie symmetry of the free ($1+2$)-dimensional L\'evy-Leblond equation}}}
\par~\par
We present, for completeness, the construction of the ${\mathbb Z}_2\times {\mathbb Z}_2$-graded Lie superalgebra symmetry of the L\'evy-Leblond equation  associated with the free ($1+2$)-dimensional heat equation.\par
 For this case the L\'evy-Leblond operator is realized by $4\times 4$ matrices. Five $4\times 4$ gamma-matrices ${\gamma_\mu}$ ($\mu=1,2,\ldots,5$), satisfying $\{\gamma_\mu,\gamma_\nu\}= 2\eta_{\mu\nu}{\mathbb I}$, with $\eta_{\mu\nu}$ a diagonal matrix with diagonal entries $(+1,+1,+1,-1,-1)$, can be introduced. We have
$\gamma_1^2=\gamma_2^2=\gamma_3^2=-\gamma_4^2=-\gamma_5^2= {\mathbb I}$. We also set
$\gamma_\pm = \frac{1}{2}(\gamma_3\pm\gamma_4)$. Expressed in the notations introduced in 
Section {\bf 2}, an explicit representation is given, e.g., by $\gamma_1= YX, \gamma_2=YX, \gamma_3= XI, \gamma_4=AI,\gamma_5= YA$.\par
The ($1+2$)-dimensional free L\'evy-Leblond equation, defined by the operator $\Omega$, is given by
\bea\label{d2levyl}
\Omega\Psi(x_1,x_2,t) =0,& & \Omega=\gamma_+\partial_t+\lambda\gamma_-+\gamma_i\partial_i.
\eea
In the above formula $i=1,2$, the Einstein convention over repeated indices is understood and $\partial_i=\partial_{x_i}$. In the following we also make use of the antisymmetric tensor $\epsilon_{ij}$, with the normalization convention $\epsilon_{12}=-\epsilon_{21}=1$.\par
The free heat equation reads
\bea\label{d2heat}
{\Omega^2}\Psi(x_1,x_2,t) = (\lambda\partial_t+\partial_i^2)\Psi(x_1,x_2,t)=0.
\eea
The following list of first-order differential symmetry operators, satisfying the (\ref{comm}) commutation property with $\Omega$, is easily derived
\bea\label{d2op1}
H&=&{\mathbb I}\partial_t,\nonumber\\
D&=& -{\mathbb I}(t\partial_t+\frac{1}{2}x_i\partial_i+\frac{1}{2})-\frac{1}{2}\gamma_+\gamma_-,\nonumber\\
K&=& -{\mathbb I}(t^2\partial_t+tx_i\partial_i-\frac{\lambda}{4}x_i^2+t) -\frac{1}{2}x_i\gamma_+\gamma_i-t\gamma_+\gamma_-,\nonumber\\
P_{+i} &=&{\mathbb I}\partial_i,\nonumber\\
P_{-i} &=& {\mathbb I}(t\partial_i-\frac{1}{2}\lambda x_i)+\frac{1}{2}\gamma_+\gamma_i,\nonumber\\
J&=& \epsilon_{ij}({\mathbb I}x_i\partial_j+\frac{1}{8}[\gamma_i,\gamma_j]),\nonumber\\
{\widetilde X}&=& \epsilon_{ij}(\gamma_+\gamma_i\partial_j+\frac{\lambda}{8}[\gamma_i,\gamma_j]),\nonumber\\
C&=&{\mathbb I}.
\eea
$D,K$ are the only operators of the above list which do not commute with $\Omega$ ($[D,\Omega]=\frac{1}{2}\Omega$, $[K,\Omega]= t\Omega$). \par
$D$ is the scaling operator. For a given operator $Z$ its scaling dimension $z$ is defined via the commutator $[D, Z]= zZ$. The scaling dimensions of the above operators are given by $z(H)=+1, z(K)=-1, z(P_{\pm i}) =\pm\frac{1}{2}$, $z(D)=z(J)=z({\widetilde X} )=z(C)=0$.\par
The operators $D,H,K$ close a $sl(2)$ symmetry algebra; the operators $P_{\pm i}, C$ close the $2$-dimensional Heisenberg-Lie algebra $h(2)$.  $J$ is the generator of the $SO(2)$ rotational symmetry.
These $3+5+1=9$ operators span the $2$-dimensional Schr\"odinger symmetry algebra $sch(2)$.\par
The physical interpretation of the remaining operator ${\widetilde X}$  entering (\ref{d2op1}) is better grasped when introducing the ${\mathbb Z}_2\times {\mathbb Z}_2$-graded extension.
For that we need to introduce symmetry operators satisfying the (\ref{anticomm}) anticommutation property with $\Omega$. We get
\bea\label{d2op2}
Q_+ &=& \gamma_+\partial_t-\lambda\gamma_-,\nonumber\\
Q_-&=& \gamma_+(t\partial_t+x_i\partial_i+1)-\frac{1}{2}\lambda x_i\gamma_i-\lambda t\gamma_-,\nonumber\\
X_i&=& \gamma_+\partial_i-\frac{1}{2}\lambda\gamma_i.
\eea
The operators $Q_+$, $X_i$ anticommute with $\Omega$ ($\{Q_+,\Omega\}=\{X_i,\Omega\}=0)$, while
$\{Q_-,\Omega\}=-\gamma_+\Omega$.\par
Their scaling dimension (measured by $D$) are $z(Q_\pm)=\pm\frac{1}{2}$, $z(X_i)$=0.\par
The operators $Q_\pm$ belong to the odd sector of a $osp(1|2)$ symmetry superalgebra whose even sector is the $sl(2)$ algebra spanned by $H,D,K$. We get, indeed,
\bea
&\{Q_+,Q_+\}= -2\lambda  H,\quad\{Q_+,Q_-\}= 2\lambda D, \quad \{Q_-,Q_-\}= 2\lambda K.&
\eea
The closure of the full set of $osp(1|2)$ anticommutation relations follows immediately.\par
We obtain, as symmetry Lie superalgebra for $\Omega$, the ${\cal N}=1$ supersymmetric extension $ssch(2)$ of the two-dimensional Schr\"odinger algebra by adding, to the
even generators spanning $sch(2)$, an odd sector containing the operators $Q_\pm$. This requires, for consistency, to add the operators $X_i$ to the odd sector, since they are recovered from the commutators
\bea
[P_{\pm i}, Q_\mp] &=& \pm X_i.
\eea
The $ {\cal N}=1$ Schr\"odinger symmetry superalgebra $ssch(2)$ is decomposed according to 
\bea
ssch(2)&=&ssch(2)_0\oplus ssch(2)_1,\nonumber\\
ssch(2)_0 &=& \{H,D,K, P_{\pm i}, C, J\},\nonumber\\
ssch(2)_1&=& \{Q_\pm, X_i\}.
\eea
It is a straightforward exercise to check the closure of the ${\mathbb Z}_2$-graded (anti)commutation relations derived for the operators entering $ssch(2)$.\par
One should note that the operator ${\widetilde X}$ entering (\ref{d2op1}) is still unaccounted for. 
The operator ${\widetilde X}$, on the other hand, naturally appears in the ${\mathbb Z}_2\times{\mathbb Z}_2$-graded extension, being obtained from the $X_i$'s commutators; we have indeed
\bea
[X_i,X_j]&=& \lambda\epsilon_{ij}{\widetilde X}.
\eea
We recall that, in the ${\mathbb Z}_2$-graded $ssch(2)$ algebra, the $X_i$'s brackets are given by the anticommutators 
\bea
\{X_i,X_j\}&=&\frac{1}{2}\lambda^2\delta_{ij} C.
\eea 
We therefore see that the introduction of the ${\mathbb Z}_2\times {\mathbb Z}_2$-graded structure offers a proper interpretation to all first-order differential operators entering (\ref{d2op1}) and (\ref{d2op2}). \par
The construction of the ${\mathbb Z}_2\times {\mathbb Z}_2$-graded Lie superalgebra goes as follows.
It requires the introduction of extra first-order and second-order differential operators $P_{s,ij}$ (with $s=\pm1, 0$) and
${\overline X}_{\pm ij}$, introduced via the anticommutators
\bea
&P_{1,ij}=\{P_{+i},P_{+ j}\}, \quad 
P_{0,ij}=\{P_{+ i},P_{- j}\}, \quad 
P_{-1,ij}=\{P_{-i},P_{- j}\} &
\eea
and
\bea
{\overline X}_{\pm, ij} &=& \{P_{\pm i}, X_j\}.
\eea
Their scaling dimensions are, respectively, $z(P_{s,ij})=s$ and $z({\overline X}_{\pm , ij}) = \pm \frac{1}{2}$.\par
We observe that the anticommutators $\{Q_\pm, X_i\}=-\lambda P_{\pm i}$ produce the first-order differential operators $P_{\pm i}$. New first-order differential operators, not entering (\ref{d2op1}) and (\ref{d2op2}),  are also obtained. We have, for instance,
\bea
\relax [Q_+, {\widetilde X}]&=& \{P_{+2},X_1\}-\{P_{+1},X_2\}= {\overline X}_{+,21}-{\overline X}_{+,12}=\lambda (\gamma_2\partial_1-\gamma_1\partial_2).
\eea
\par
The ${\mathbb Z}_2\times {\mathbb Z}_2$-graded symmetry structure is introduced as a ${\mathbb Z}_2\times {\mathbb Z}_2$-graded Lie superalgebra, with brackets, the (anti)commutators, defined in (\ref{brackets}) and based on the (\ref{z2superalg}) inner product.\par
The vector space of the ${\mathbb Z}_2\times {\mathbb Z}_2$-graded Lie superalgebra ${\cal L}_{{\mathbb Z}_2\times{\mathbb Z}_2}$ is decomposed according to
\bea
{\cal L}_{{\mathbb Z}_2\times{\mathbb Z}_2}&=& {\cal L}_{00}\oplus {\cal L}_{01}\oplus {\cal L}_{10}\oplus
{\cal L}_{11},
\eea
with the different sectors respectively spanned by the operators
\bea\label{z2z2d2}
{\cal L}_{00} &=& \{H,D,K, J, {\widetilde X}, P_{s,ij}\},\nonumber\\
{\cal L}_{01} &=& \{ P_{\pm i}\},\nonumber\\
{\cal L}_{10}&=& \{ Q_\pm, {\overline X}_{\pm, ij}\},\nonumber\\
{\cal L}_{11} &=& \{ X_i\}. 
\eea
With lenghty and tedious, but straighforward, computations the following three main properties are proven:\\
{\em i}) all operators in (\ref{z2z2d2}), including the second-order differential operators, are symmetry operators, satisfying either the (\ref{comm}) or the (\ref{anticomm}) conditions with respect to  the operator $\Omega$ given in (\ref{d2levyl}),\\
{\em ii}) the (\ref{brackets},\ref{z2superalg}) brackets, defined by the respective (anti)commutators, are closed on the (\ref{z2z2d2}) generators and, finally,\\
{\em iii}) the ${\mathbb Z}_2\times {\mathbb Z}_2$-graded Jacobi identities (\ref{gradedjac}) are satisfied by the (\ref{z2z2d2}) generators.\par
 These three main results prove that ${\cal L}_{{\mathbb Z}_2\times {\mathbb Z}_2}$ is a ${\mathbb Z}_2\times{\mathbb Z}_2$-graded Lie superalgebra symmetry of the L\'evy-Leblond square root of the ($1+2$)-dimensional free heat equation.\par
By adding the identity operator $C$ to the ${\cal L}_{00}$ sector of ${\cal L}_{{\mathbb Z}_2\times{\mathbb Z}_2}$
we obtain the ${\mathbb Z}_2\times{\mathbb Z}_2$-graded Lie superalgebra ${\cal L}_{{\mathbb Z}_2\times{\mathbb Z}_2}'$, given by
\bea\label{newz2z2}
 {\cal L}_{{\mathbb Z}_2\times{\mathbb Z}_2}' &=&{\cal L}_{{\mathbb Z}_2\times{\mathbb Z}_2}\oplus u(1).
\eea
%
%
~
\\ {~}~{~}
\\ {~}~{~}
\renewcommand{\theequation}{C.\arabic{equation}}
\setcounter{equation}{0}

{\Large{\bf Appendix C: On different grading assignments for the symmetry operators}}
\par~\par
In Appendix {\bf A} we showed, with examples taken from quaternions and split-quaternions, that different ${\mathbb Z}_2\times {\mathbb Z}_2$-graded Lie (super)algebras can be defined on the same vector space,  
based on a different grading assignment for the same given set of operators. \par
It is tempting to investigate this feature for the symmetry operators of the free L\'evy-Leblond equation. Since, in particular, the natural finite ${\mathbb Z}_2\times {\mathbb Z}_2$-graded Lie superalgebra (\ref{z2z2sym}) is induced from the repeated (anti)commutators of the operators $P_{\pm\frac{1}{2}}\in {\cal G}_{01}$, $Q_{\pm \frac{1}{2}}\in{\cal G}_{10}$, one can ask which is the output if $P_{\pm\frac{1}{2}}$ (given in (\ref{sigmasym},\ref{equivalPP})) are kept in the $ {\cal G}_{01}$ sector, while $Q_{\pm \frac{1}{2}}$ (given in (\ref{qplus},\ref{qminus})) are assigned to the ${\cal G}_{11}$ sector of a ${\mathbb Z}_2\times {\mathbb Z}_2$-graded Lie superalgebra. The consistency condition requires to derive the operator $X\in {\cal G}_{10}$ from the commutators $[\pm P_\pm, Q_\pm]=X$. The commutator 
$[Q_+,Q_-]={\overline Q}_0$ produces the first-order differential operator ${\overline Q}_0\in {\cal G}_{00}$.
The further commutators $\relax [P_{\pm\frac{1}{2}},{\overline Q}_0]$ produce the extra (\ref{secondosp}) pair of $\Omega_{\pm\frac{1}{2}}$ operators. One should note that $\Omega_{\pm\frac{1}{2}}$ are assigned to ${\cal G}_{01}$ and that $\Omega_{+\frac{1}{2}}$ is the original L\'evy-Leblond operator introduced in equation (\ref{omega}).\par
The construction has to be continued. The new grading assignment produces a ${\mathbb Z}_2\times {\mathbb Z}_2$-graded Lie superalgebra which, unlike (\ref{z2z2sym}), is infinite-dimensional and spanned by differential operators of any order. To check this statement it is sufficient to compute the repeated commutators
$[Q_{+\frac{1}{2}},[Q_{+\frac{1}{2}},\ldots[Q_{+\frac{1}{2}},{\overline Q}_0]\ldots]]$.   
\par
For completeness we produce here a finite ${\mathbb Z}_2\times{\mathbb Z}_2$-graded Lie superalgebra, defined abstractly in terms of (anti)commutators and whose consistency is implied by the fact that the graded Jacobi identities are satisfied.  This algebra cannot be realized by the L\'evy-Leblond symmetry operators. 
It shares, nevertheless, some features in common. We impose, in particular, the existence of two $osp(1|2)$ subalgebras, whose generators (as in Section {\bf 5}) are expressed as
$ \{ P_0, P_{\pm \frac{1}{2}}, P_{\pm 1} \} $ and $ \{ H, D, K, Q_{\pm \frac{1}{2}} \} $. Their respective (anti)commutators are given by equations (\ref{schr},\ref{posp},\ref{dosp},\ref{qosp},\ref{hkosp}). Equation (\ref{pqx}) introduces the extra generator $X$. An extra set of generators, $R_0,R_{\pm 1}, Y, Z, C$, are introduced through the positions
\begin{equation}
 R_0 = [Q_{\frac{1}{2}}, Q_{-\frac{1}{2}}], \quad 
 R_{\pm 1} = \{ Q_{\pm{\frac{1}{2}}}, P_{\pm{\frac{1}{2}}} \}, \quad 
 Y = \{ Q_{-\frac{1}{2}}, P_{\frac{1}{2}} \}, \quad 
 Z = \{ Q_{\frac{1}{2}}, P_{-\frac{1}{2}} \},
\end{equation}
while $C$ is the center of the algebra.\par
A consistent ${\mathbb Z}_2\times{\mathbb Z}_2$-graded Lie superalgebra assignment is realized by spanning the graded vector spaces as follows:
\bea\label{z2z2sym2}
{\cal G}_{00} &=& \{ H, D, K, P_{\pm 1}, P_0, R_0, C \},\nonumber\\
{\cal G}_{01} &=& \{ P_{\pm \frac{1}{2}}\},\nonumber\\
{\cal G}_{10} &=& \{ X, R_{\pm 1}, Y, Z \},\nonumber\\
{\cal G}_{11} &=& \{ Q_{\pm{\frac{1}{2}}} \}.
\eea 
The remaining non-vanishing (anti)commutators which define the  ${\mathbb Z}_2\times{\mathbb Z}_2$-graded Lie superalgebra structure of (\ref{z2z2sym2}) are given by
\begin{equation}
 \begin{array}{lcl}
   [H, Y] = R_1, & & [H, Z] = R_1 \\[5pt]
   [H, R_{-1}] = Y+Z, & & [D, R_{\pm 1}] = \pm R_{\pm 1}, \\[5pt]
   [K, R_1] = Y+Z, & & [K, Y]=R_{-1}, \\[5pt]
   [K, Z] = R_{-1}, & & [R_0, X] = Y-Z, \\[5pt]
   \{X, X\} = C, & & \{X, R_{\pm 1} \} = - P_{\pm 1}, \\[5pt]
   \{X, Y \} = -P_0, & & \{ X, Z \} = -P_0, \\[5pt]
   \{X, Q_{\pm \frac{1}{2}} \} = -P_{\pm \frac{1}{2}}. & & 
 \end{array}
\end{equation}
These (anti)commutators guarantee the closure of the graded Jacobi identities for (\ref{z2z2sym2}).\par
Unlike (\ref{z2z2sym}), the finite ${\mathbb Z}_2\times{\mathbb Z}_2$-graded Lie superalgebra  (\ref{z2z2sym2}) is not a symmetry algebra of the L\'evy-Leblond equation. It is an open question whether it appears as the symmetry algebra of some dynamical system.

\end{document}